\documentstyle[11pt,psfig]{article}

\newtheorem{theorem}{Theorem}

\title{ A Method to Extract Potentials from the Temperature
Dependence of Langmuir Constants for Clathrate-Hydrates }

\author{Martin Z. Bazant \\
{\small {\it Department of Mathematics, Massachusetts
Institute of Technology, Cambridge, MA 02139}} \\
\ \\
Bernhardt L. Trout
\\
{\small {\it Department of Chemical Engineering, Massachusetts
Institute of Technology, Cambridge, MA 02139}}
}

\date{July 7, 2001}

\begin{document}

\maketitle

\begin{abstract}


It is shown that the temperature dependence of Langmuir constants
contains all the information needed to determine spherically averaged
intermolecular potentials. An analytical ``inversion'' method based on
the standard statistical model of van der Waals and Platteeuw is
presented which extracts cell potentials directly from experimental
data. The method is applied to ethane and cyclopropane
clathrate-hydrates, and the resulting potentials are much simpler and more
meaningful than those obtained by the usual method of numerical
fitting with Kihara potentials.

\end{abstract}






\section{Introduction}

A central mission of chemical physics is to determine intermolecular
interactions from experimental phase equilibrium data.  This is
generally a very difficult task because macroscopic equilibrium
constants reflect averaging over vast numbers of poorly understood
microscopic degrees of freedom. Since the advent of the computer,
empirically guided numerical fitting has become the standard method to
obtain intermolecular potentials. An appealing and often overlooked
alternative, however, is to solve {\it inverse problems} based on
simple statistical mechanical models.

A well-known class of inverse problems is to determine the density of
states $n(E)$ from the partition function $Q(\beta)$, or one of its
derivatives, as a function of inverse temperature $\beta = 1/kT$. For
classical Maxwell-Boltzmann statistics, these quantities are simply
related by an Laplace transform~\cite{pathria}
\begin{equation}
Q(\beta) = \int_0^\infty e^{-\beta \varepsilon} n(\varepsilon) d\varepsilon
\label{eq:QMB}
\end{equation}
which is straightforward to invert. Analogous quantum-mechanical
problems have also been solved using Laplace (or Mellin) transform
methods. For example, in Bose-Einstein statistics,
\begin{equation}
Q(\beta) = \int_0^\infty \frac{n(\varepsilon) d\varepsilon}{e^{\beta
\varepsilon} + 1},
\end{equation}
the phonon density of states can be obtained from the specific heat of
a
crystal~\cite{weiss,pindor,igalson,chen,hughes,pincomment,chenphonon,ninham},
and the area distribution of a blackbody radiator can be obtained from
its power spectrum~\cite{lak,boj,kim,chen,hughes}. Similarly, in
Fermi-Dirac statistics,
\begin{equation}
Q(\beta) = \int_0^\infty \frac{n(\varepsilon) d\varepsilon}{e^{\beta
(\varepsilon-\varepsilon_F)} - 1},
\end{equation}
the band structure of a doped semiconductor can be obtained from the
temperature dependence of its carrier density~\cite{chenfermi}. In
each of these examples, it is possible to extract the microscopic
density of states from a temperature-dependent thermodynamic quantity
because each is a function of just one variable.

In the more complicated case of chemical systems, thermodynamic
quantities are related to classical configurational integrals,
\begin{equation}
Z(\beta) = \int_V e^{-\beta \Phi(\vec{r})} d\vec{r},
\end{equation}
where the total energy $\varepsilon$ in Eq.~(\ref{eq:QMB}) is replaced
by the intermolecular potential $\Phi(\vec{r})$ , and the integral
extends over the interaction volume $V$. Unfortunately, these
integrals are multi-dimensional, which leads to under-determined
inverse problems. Perhaps as a result, statistical inversion methods
have apparently not been developed for intermolecular potentials in
chemical systems, and instead empirical fitting has been used
exclusively to describe phase equilibrium data.

In this article, we show that within the common spherical-cell
approximation {\it the intermolecular potential is completely
determined by the temperature dependence of the Langmuir constant.}
In this case, the linear integral equation (\ref{eq:QMB}) is replaced
by a non-linear equation of the form
\begin{equation}
Z(\beta) = 4\pi \int_0^\infty e^{-\beta w(r)} r^2 dr
\label{eq:Finteqn}
\end{equation}
where $w(r)$ is a spherically averaged ``cell
potential''~\cite{ljd}. As shown below in the important case of
Langmuir constants for clathrate-hydrates, this simple inverse problem
can be solved exactly without resorting to numerical fitting schemes.

Before proceeding, we mention some related ideas in the recent
literature of solid state physics.  As with chemical systems,
empirical fitting is also the standard approach to derive interatomic
potentials for metals and semiconductors. Since the pioneering work of
Carlsson, Gelatt, and Ehrenreich in 1980~\cite{cge,carlsson}, however,
exact inversion methods have been developed to obtain potentials from
cohesive energy curves~\cite{chen,chenren,chen94,bk1,bk2,edip}.  These
theoretical advances, discussed briefly in Appendix A, have recently
led to improvements in the modeling of silicon, beyond what had
previously been obtained by empirical fitting alone~\cite{edip,justo}.
Inspired by such developments in solid state physics, here we seek
similar insights into clathrate-hydrate intermolecular forces, albeit
using a very different statistical mechanical formalism.

\section{ The Inverse Problem for Clathrate-Hydrates }

\subsection{ The Statistical Theory of van der Waals and Platteeuw}

Clathrate-hydrates exist throughout nature and are potentially very
useful technological materials~\cite{sloan}. For example, existing
methane hydrates are believed to hold much more energy than any fossil
fuel in use today. Carbon dioxide hydrates are being considered as
effective materials for the sequestration and/or storage of CO$_{2}$.
In spite of their great importance, however, the theory of
clathrate-hydrate phase behavior is not very well developed, still
relying for the most part on the {\it ad hoc} empirical fitting of
experimental data. Therefore, we have chosen to develop our
statistical inversion method in the specific context of
clathrate-hydrate chemistry.

Since being introduced in 1959, the statistical thermodynamical model
of van der Waals and Platteeuw (vdWP) has been used almost exclusively
to model the phase behavior of clathrate-hydrates, usually together
with a spherical cell (SC) model for the interaction potential between
the enclathrated or ``guest'' molecule and the cage of the
clathrate-hydrate.  The SC model was also introduced by vdWP, inspired
by an analogous approximation made by Lennard-Jones and Devonshire in
the case of liquids~\cite{vdwp,ljd}.

In the general formulation of vdWP~\cite{vdwp}, the chemical potential
difference between an empty, unstable hydrate structure with no guest
molecules, labeled MT, and the stable hydrate, labeled H, is related
to the so-called Langmuir hydrate constant $C_{Ji}$ and the fugacity of the
guest molecule $\hat{f}_J$
\begin{equation}
\Delta\mu^{MT-H} = kT\sum_{i}\nu_{i}\ln(1+\sum_{J}C_{Ji}\hat{f}_J)
\label{eq:vdw-p}
\end{equation}
where $i$ designates the type of cage, $\nu_{i}$ the number of cages
of type $i$ per water molecule and $J$ the type of guest molecule. In
practice, experimental phase equilibria data is used to
determine $\Delta\mu^{MT-H}$.

The connection with intermolecular forces within vdWP theory is
made by expressing the Langmuir hydrate constant as the configurational
integral $Z_{Ji}$ divided by $kT$, which is written explicitly as an
integral over the volume $V$
\begin{equation}
C_{Ji}(T) = \frac{1}{8\pi^{2}kT}\int_V e^{-
\Phi(r,\theta,\phi,\alpha,\xi,\gamma)/kT} r^{2} sin \theta sin \xi dr
d\theta d\phi d\alpha d\xi d\gamma
\label{eq:Phi}
\end{equation}
where $\Phi(r,\theta,\phi,\alpha,\xi,\gamma)$ is the general six
dimensional form of the interaction potential between the guest
molecule at spherical coordinates $(r,\theta,\phi)$ oriented with
Euler angles $(\alpha,\xi,\gamma)$ with respect to all of the water
molecules in the clathrate-hydrate.

In the SC approximation, which is made without any careful
mathematical justification, the intermolecular potential $\Phi$ is
replaced by a spherically averaged cell potential $w(r)$, which
reduces the Langmuir constant formula (\ref{eq:Phi}) to a single,
radial integration
\begin{equation}
C_{Ji}(T) = \frac{4\pi}{kT} \int_0^R e^{-w(r)/kT}r^2 dr
\label{eq:C_LJD}
\end{equation}
where the cutoff distance $R$ is often arbitrarily taken as the radius
of the cage. (As shown below, the exact value rarely matters because
temperatures are typically so low that the high energy portion of the
cage $r \approx R$ makes a negligible contribution to the integral.)
Although the SC approximation may appear to be a drastic
simplification, it is nevertheless very useful for theoretical studies
of intermolecular forces based on Langmuir constant measurements.

\subsection{ Numerical Fitting Schemes}

Before this work, the functional form of the cell potential $w(r)$ has
always been obtained by first choosing a model interaction potential
between the guest molecule in a cage and each nearest neighbor water
molecule essentially {\it ad hoc}, and then performing the spherical
average from (\ref{eq:Phi}) to (\ref{eq:C_LJD}) analytically.  The
most common potential form in use today is the Kihara potential, which
is simply a shifted Lennard-Jones potential with a hard-core. Using
the Kihara potential and spherically averaging the interaction energy,
typically over the first-shell only, yields the following functional
form for $w(r)$:
\begin{equation}
w(r) = 2z\epsilon[\frac{\sigma^{12}}{R^{11}r}(\delta_{10} +
\frac{a}{R}\delta_{11}) - \frac{\sigma^{6}}{R^{5}r}(\delta_{4} +
\frac{a}{R}\delta_{5})]
\label{eq:w-Kihara}
\end{equation}
where
\begin{equation}
\delta_{N}=\frac{1}{N}[(1-\frac{r}{R}-\frac{a}{R})^{-N} -
(1+\frac{r}{R}-\frac{a}{R})^{-N}]
\label{eq:delta}
\end{equation}
and $z$ is the coordination number, $R$ is the radius of the cage, and
$\sigma$, $\epsilon$, and $a$ are the Kihara parameters.  As a result
of the averaging process leading from (\ref{eq:Phi}) to
(\ref{eq:C_LJD}), the functional form of $w(r)$ is fairly complicated,
and the parameters $\epsilon$ and $\sigma$ are generally determined by
fitting monovarient equilibrium temperature-pressure data numerically
\cite{sloan,pp}.

There are several serious drawbacks to this ubiquitous numerical
fitting procedure, which suggest that the Kihara parameters lack any
physical significance: $(i)$ The Kihara parameters are not unique, and
many different sets can fit the experimental data well; $(ii)$ the
Kihara parameters found by fitting Langmuir curves do not match those
of found by fitting other experimental data, such as the second virial
coefficient or the gas viscosity~\cite{sloan}; and $(iii)$ comparisons
of Langmuir constants found via the SC approximation (\ref{eq:C_LJD})
and via explicit multi-dimensional quadrature (\ref{eq:Phi}) show that
the two can differ by over 12 orders of
magnitude~\cite{holder,sparks_cao} (which results from the
exponentially strong sensitivity of the Langmuir constant to changes
in the cell potential).  These problems call into question the
validity of using the Kihara potential as the basis for the empirical
fitting, and even the use of the SC approximation itself.

\subsection{ Inversion of Langmuir Curves }

\begin{figure}
\begin{center}
\mbox{
\psfig{file=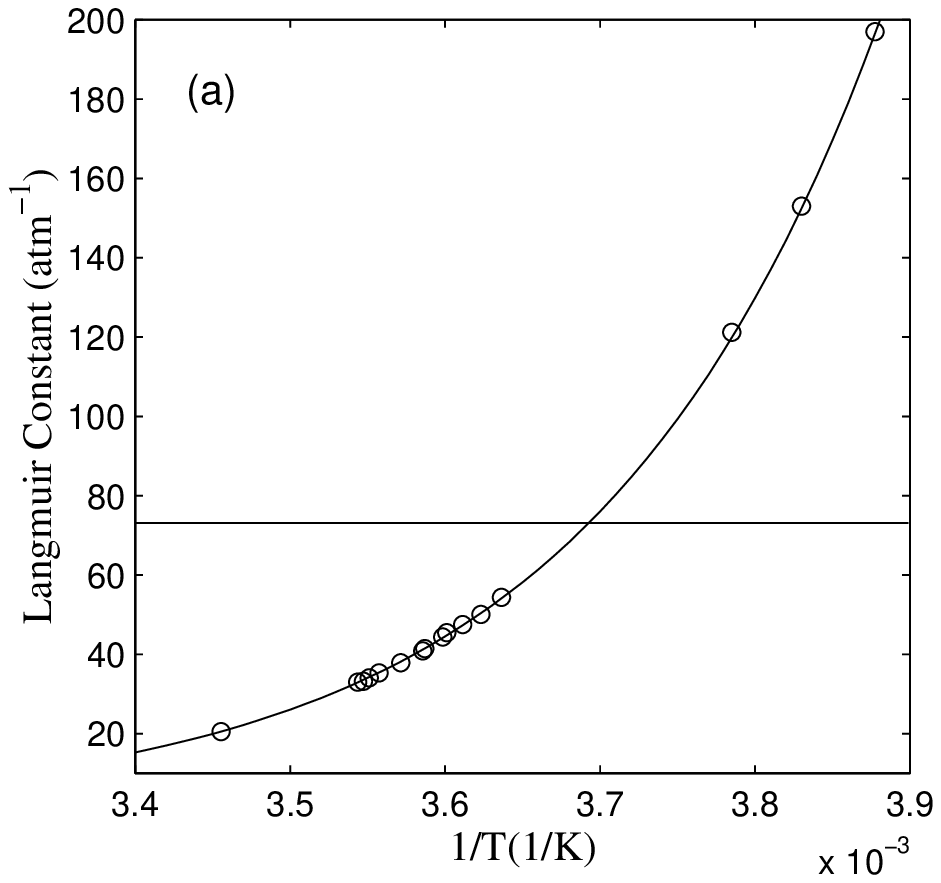,height=2.4in}
} \\
\vspace{0.2in}
\mbox{
\psfig{file=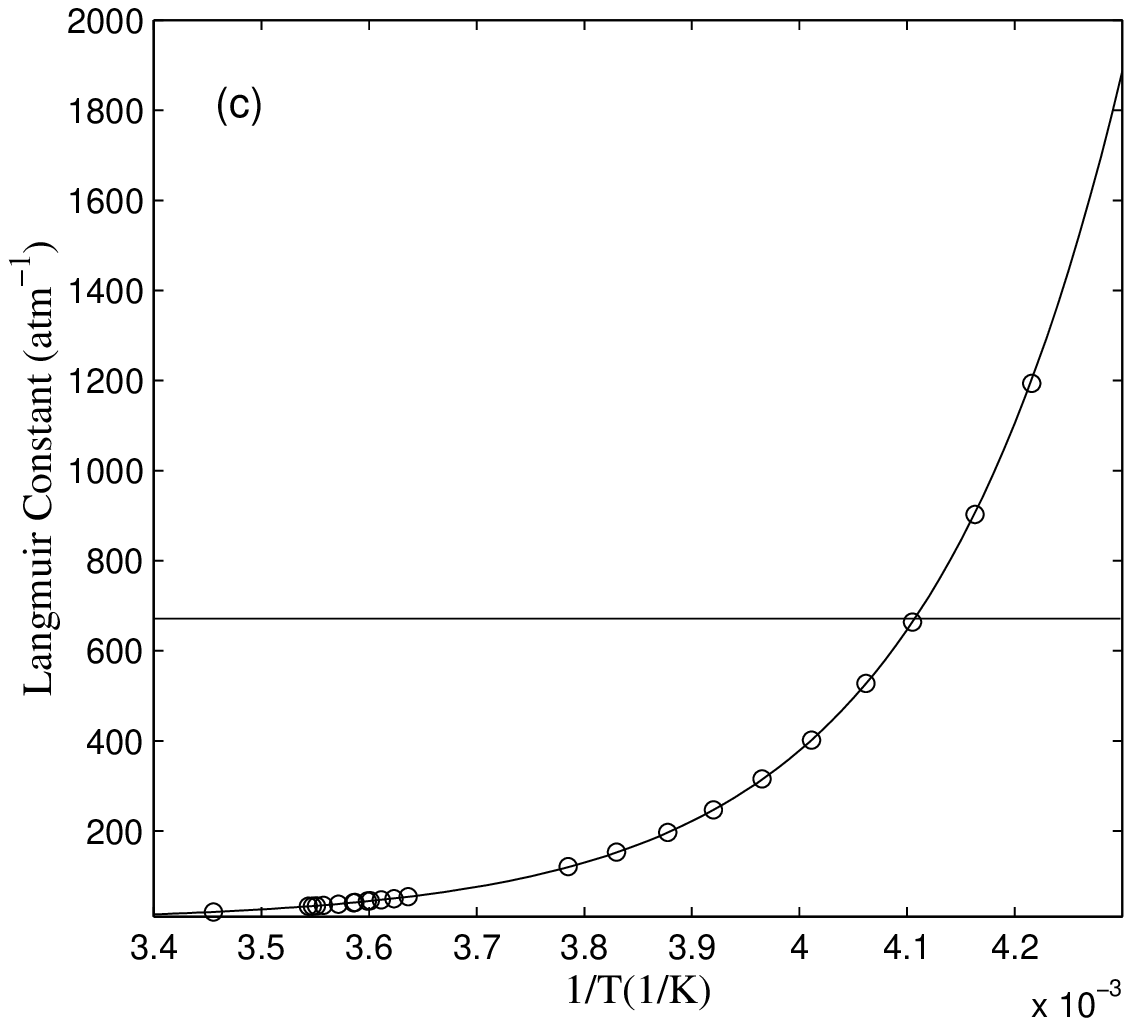,height=2.4in}
} \\
\vspace{0.2in}
\mbox{
\psfig{file=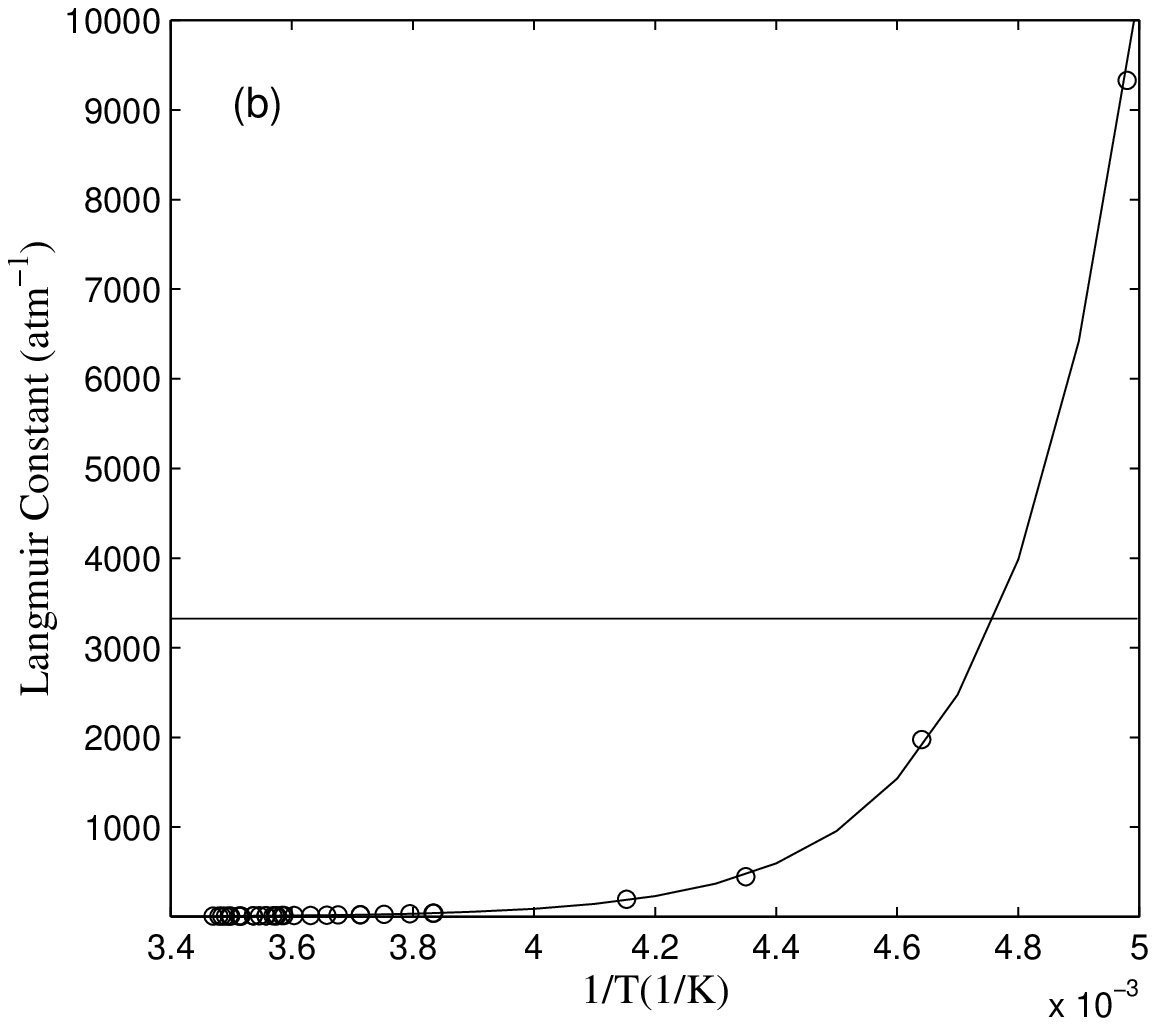,height=2.4in}
} \\
\vspace{0.2in}
\begin{minipage}[h]{5in}
\caption{ 
Exponential fits of Langmuir constants over the measured
temperature range plotted with linear axes for (a)-(b) cyclopropane
and (c) ethane clathrate-hydrates. An enlargement of the high
temperature data for cyclopropane is shown in (a). The experimental
data is taken from Ref.~\protect\cite{sparks_thesis}.
\label{fig:Cexpt-linear}
}
\end{minipage}
\end{center}
\end{figure}

It would clearly be preferable to extract more reliable information
about the interatomic forces in clathrate-hydrates directly from
experimental data without any {\it ad hoc} assumptions about their
functional form. In principle, such an approach is possible for
clathrate-hydrates which contain a single type of guest molecule
occupying only one type of cage.  In this case, each of the sums in
Eq.~(\ref{eq:vdw-p}) contains only one term, and by using an equation
of state to compute the fugacity $\hat{f}_J$, the Langmuir constant
$C_{Ji}$ can be determined directly from experimental phase equilibria
data.  Typical data sets obtained in this manner are shown in
Fig.~\ref{fig:Cexpt-linear} for Structure I ethane and cyclopropane
clathrate-hydrates~\cite{sparks_thesis}.

Because the full potential $\Phi$ in (\ref{eq:Phi}) is
multi-dimensional (while the Langmuir constant only depends on a
single parameter $T$), the general vdWP theory is too complex to pose
a well-defined inverse problem for the interatomic forces. The SC
approximation, on the other hand, introduces a very convenient
theoretical construct, the spherically averaged potential $w(r)$,
which has the same dimensionality as the Langmuir curve $C_{Ji}(T)$ of
a single type of guest molecule occupying a single type of
cage. (Since we consider only this case, we drop the subscripts $Ji$
hereafter.)  Although one can question the accuracy of the SC
approximation, its simplicity at least allows precise connections to
be made between the Langmuir curve and the cell potential.

As an appealing alternative to empirical fitting, in this article we
view Eq.~(\ref{eq:C_LJD}) as an integral equation to be solved
analytically for $w(r)$, given a particular Langmuir curve
$C(T)$. Letting $\beta = 1/kT$, we rewrite (\ref{eq:C_LJD}) as
\begin{equation}
C(\beta) = 4\pi\beta \int_0^\infty e^{-\beta w(r)}r^2 dr,
\label{eq:langmuir}
\end{equation}
where we have also set the upper limit of integration to $R=\infty$,
which introduces negligible errors due to the very low temperatures
(large $\beta$) accessible in experiments. (This will be justified
{\it a posteriori} with a precise definition of ``low'' temperatures
below.)  Note that Eq.~(\ref{eq:langmuir}) has the form of
Eq.~(\ref{eq:Finteqn}) with $Z(\beta) = \beta C(\beta)$.

\subsection{ Application to Experimental Data}

\begin{figure}
\begin{center}
\mbox{
\psfig{file=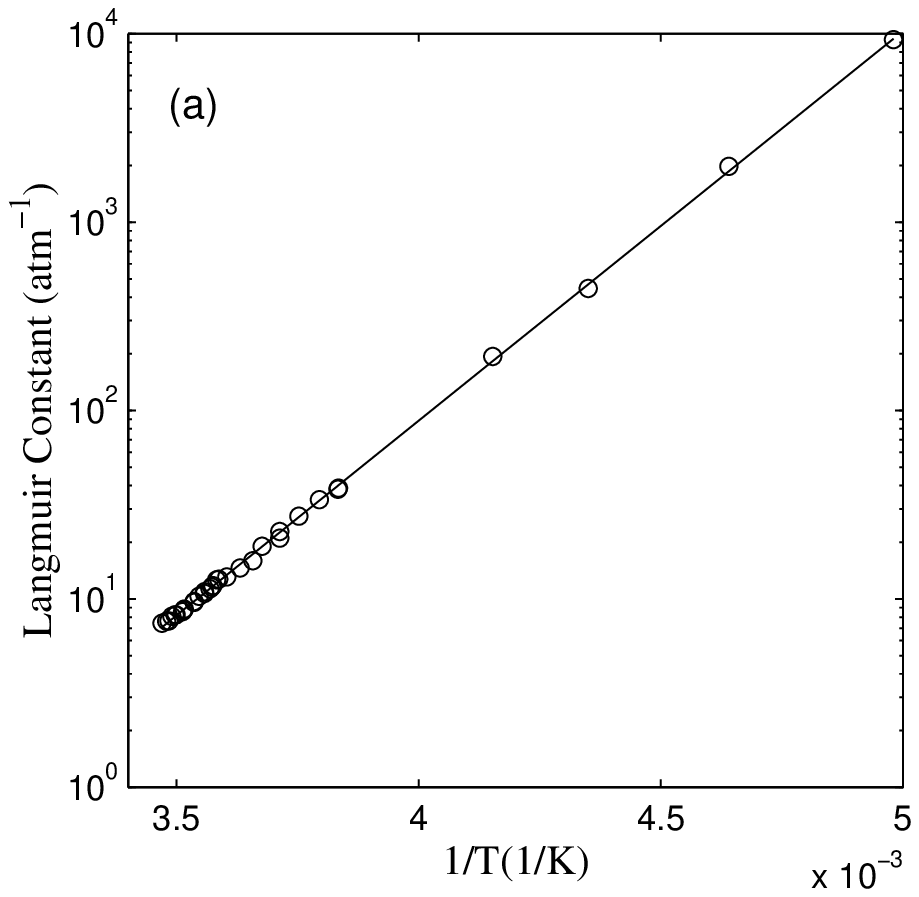,width=3.5in}
}\\
\vspace{0.2in}
\mbox{
\psfig{file=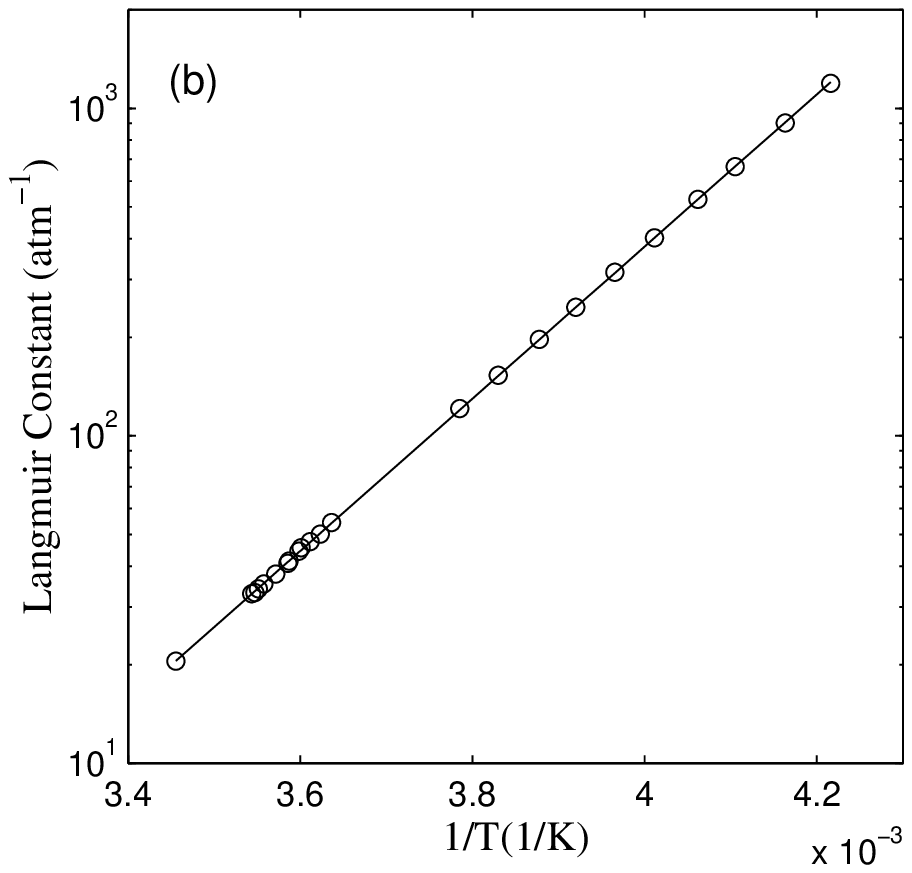,width=3.5in}
}\\
\vspace{0.2in}
\begin{minipage}[h]{5in}
\caption{
Exponential dependence with inverse temperature of
experimental Langmuir curves from Fig.~\ref{fig:Cexpt-linear} plotted
with log-linear axes for (a) ethane and (b) cyclopropane clathrate
hydrates. Straight lines indicate pure van't Hoff behavior.
\label{fig:Cexpt} }
\end{minipage}
\end{center}
\end{figure}

In our analytical approach, some straight-forward fitting of the raw
experimental data is needed to construct the function $C(\beta)$, but
after that, the ``inversion'' process leading to $w(r)$ is exact. For
example, typical sets of experimental data are well described by a
van't Hoff temperature dependence
\begin{equation}
C(\beta) = C_o e^{m\beta}
\label{eq:Cexpt}
\end{equation}
as shown in Fig.~\ref{fig:Cexpt} for ethane and cyclopropane clathrate
hydrates~\cite{sparks_thesis}, and the constant $m$ is generally
positive. (Note that the exponential dependence which we call "van't
Hoff dependence" in this paper can be expected based on quite general
thermodynamic considerations~\cite{TM}.)

Aided by the analysis, the quality and functional form of
these fits are discussed below in section~\ref{sec:expt}. In order to
allow for deviations from the dominant van't Hoff behavior, however, in
this article we consider the more general form
\begin{equation}
C(\beta) = \beta F(\beta) e^{m\beta}
\label{eq:Fdef}
\end{equation}
where $m$ is a constant defined by
\begin{equation}
m = \lim_{\beta\rightarrow\infty} \log C(\beta)/\beta
\label{eq:mdef}
\end{equation}
whenever this limit exists and is finite, i.e. when the prefactor
$F(\beta)$ in Eq.~(\ref{eq:Fdef}) is dominated by the exponential term
at low temperatures.  We exclude the possibility of hyper-exponential
behavior at low temperatures, $\log C/\beta \rightarrow \infty$, which
is not physically meaningful, as explained below.  The set of possible
prefactors includes power-laws, $F(\beta)=\beta^{-\mu}$, as well as
various rational functions.

The rest of article is organized as follows.  In section~\ref{sec:math} we
discuss various necessary and sufficient conditions for the existence
of physically reasonable solutions, and we also derive the asymptotics
of the Langmuir curve at low temperature from the behavior of the cell
potential at near its minimum. In section~\ref{sec:general}, we
perform the analysis in the general case, and in
section~\ref{sec:vanthoff}, we discuss the specific case of van't Hoff
dependence (\ref{eq:Cexpt}), which leads to a cubic solution as well
as various unphysical solutions involving cusps.  In
section~\ref{sec:deviations}, we derive analytical solutions for
several different temperature dependences, which reveal the
significance of possible deviations from van't Hoff behavior for the
form of the potential $w(r)$.  The theoretical curves are compared
with the experimental data in section~\ref{sec:expt}, and the physical
conclusions of the analysis are summarized in section~\ref{sec:concl}.
Relevant mathematical theorems are proved in the Appendix B.

\section{General Analysis of the Inverse Problem}
\label{sec:math}

\subsection{Necessary Conditions for the Existence of Solutions}

On physical grounds, it expected that the cell potential $w(r)$ is
continuous (at least piecewise) and has a finite minimum at $r_o \geq
0$ somewhere inside the clathrate cage, $w(r) \geq w(r_o) = w_o$. We also
allow the possibility that $w(r)$ is infinite for certain values of
$r$ (e.g. outside a ``hard-wall radius'') by simply omitted such
values from the integration in Eq.~(\ref{eq:langmuir}).  As proved in
the Appendix B, these simple physical requirements suffice to imply the
asymptotic relation (\ref{eq:mdef}), where $m=-w_o$. They also
place important constraints on the prefactor $F(\beta)$ defined in
(\ref{eq:Fdef}), which must be
\begin{itemize}
\item[(i)] analytic in the half plane $\mbox{Re}\beta > c$ and
\item[(ii)] real, positive and non-increasing for $\beta > c$ on the
real axis
\end{itemize}
where $c \geq 0$ is a real number.  (Note that we view the inverse
temperature $\beta$ as a complex variable, for reasons soon to become
clear.)  Moreover, if the set
\begin{equation}
S = \{ r\geq 0 | w_o < w(r) < \infty \}
\end{equation}
has nonzero measure, then $F(\beta)$ is strictly decreasing on the
positive real axis. It is straightforward to generalize these rigorous
results to the multi-dimensional integral of vdWP theory,
Eq.~(\ref{eq:Phi}), without making the spherical cell approximation,
as described in the Appendix B, but hereafter we discuss only the
spherically averaged integral equation, Eq.~(\ref{eq:langmuir}),
because it makes possible an exact inversion.

In this section, we give simple arguments to explain the results
proved in Theorem 1 of the Appendix B.  First, we consider the
illustrative example of a constant cell potential with a hard wall at
$r=r_{hw}>0$,
\begin{equation}
w(r) = \left\{ \begin{array}{ll}
w_o & \mbox{ if } 0 \leq r < r_{hw} \\
\infty &  \mbox{ if } r > r_{hw}
\end{array}
\right.
\end{equation}
which satisfies the assumptions stated above.  The integral
(\ref{eq:langmuir}) is easily performed in this case to yield
\begin{equation}
C(\beta) = \frac{4}{3}\pi r_{hw}^3 \beta e^{-w_o \beta}
\end{equation}
which implies $m=-w_o$, the well depth, and $F(\beta) = \frac{4}{3}\pi
r_{hw}^3$, the volume of negative energy. Consistent with the general
results above, $F(\beta)$ is constant in this case since
$S=\emptyset$. For continuous potentials $w(r)$, however, the prefactor
$F(\beta)$ must be strictly decreasing because $S \neq \emptyset$.

The integral equation (\ref{eq:langmuir}) can be simplified by a
change of variables from radius to volume.  In terms of a shifted cell
potential versus volume
\begin{equation}
u(x) = w(r) - w_o \ \ \mbox{ where } \ \ x = \frac{4\pi}{3}r^3
\end{equation}
the integral equation is reduced to the form
\begin{equation}
C(\beta) = \beta F(\beta) e^{-w(r_o) \beta}
\end{equation}
where
\begin{equation}
F(\beta) = \int_0^\infty e^{-\beta u(x)} dx.
\label{eq:u}
\end{equation}
Since $u(x) \geq 0$ by construction, the function $F(\beta)$ is
clearly non-increasing. In the Appendix B, it is proved that if the
potential varies continuously near its minimum (in a very general
sense), then $F(\beta)$ does not decay exponentially.  Since
$F(\beta)$ also positive and bounded above, we conclude
\begin{equation}
\lim_{\beta\rightarrow\infty} \log C(\beta)/\beta = -w_o
\label{eq:logCasymp}
\end{equation}
which implies $m = -w_o$. Therefore, {\it the slope of the Langmuir
curve on a ``van't Hoff plot''} ($\log C$ versus $\beta = 1/kT$) {\it
in the low temperature limit is equal to (minus) the minimum energy of
the cell potential}. Since it is generally observed that $m$ is
positive, as in the case of ethane and cyclopropane clathrate-hydrates
shown in Figs.~\ref{fig:Cexpt-linear}--\ref{fig:Cexpt}, the cell
potential must be attractive, $w_o = -m < 0$, which simply indicates
that the total internal energy is lowered by the introduction of guest
molecules into clathrate-hydrates.

The fact that $F(\beta)$ must be nondecreasing has important
consequences for the existence of solutions which are piecewise
continuous and bounded below.  For example, consider the class of
Langmuir curves of the form
\begin{equation}
C(\beta) = \beta^\nu e^{m\beta}
\end{equation}
which is useful in fitting experimental data (see below). We have
already addressed the borderline case, $\nu = 1$, in which a
discontinuous hard-wall solution is possible. Although it is not
obvious {\it a priori}, there are no solutions to the inverse
problem if $\nu > 1$, since in that case $F(\beta) = \beta^{\nu-1}$ is
increasing. On the other hand, if $\nu < 1$, then well-behaved
continuous solutions are possible, because $F(\beta)$ is strictly
decreasing.

\subsection{Low Temperature Asymptotics of the Langmuir Curve}

>From Eq.~(\ref{eq:logCasymp}), we see that the minimum energy $w_o =
w(r_o)$ determines the leading order asymptotics of $C(\beta)$ in
the low temperature limit.  More generally, one would expect that
$C(\beta)$ at low temperatures is completely determined by shape of
the cell potential at low energies, close to its minimum. Using
standard methods for the asymptotic expansion of Laplace
integrals~\cite{bender}, it is straightforward to provide a
mathematical basis for this intuition. For simplicity, here we
consider the usual case of a parabolic minimum
\begin{equation}
w(r) = w_o + \frac{1}{2}k(r-r_o)^2 + O((r-r_o)^3)
\label{eq:wpara}
\end{equation}
for some constants $k>0$ and $r_o\geq 0$, although below we will
derive many exact solutions with non-parabolic minima.

Due to the factor of $r^2$ appearing in the integrand in
Eq.~(\ref{eq:langmuir}), the two cases of a non-central or central
minimum, $r_o > 0$ and $r_o=0$, respectively, must be treated
separately.  Physically, this qualitative difference between central
and non-central-wells is due to the spherical averaging process going
from Eq.~(\ref{eq:Phi}) to Eq.~(\ref{eq:langmuir}): A central well in
$w(r)$ corresponds to a unique local minimum of the multidimensional
potential $\Phi$, but a non-central-well in $w(r)$ corresponds to a
nonlocal minimum of $\Phi$ which is smeared across a sphere of radius
$r_o$.

Beginning with non-central-well case, $r>0$, we have the
following asymptotics as $\mbox{Re}\beta \rightarrow \infty$:
\begin{eqnarray}
C(\beta) &\sim & 4\pi r_o^2 \beta \int_{r_o-\epsilon}^{r_o+\epsilon}
e^{-\beta(w_o + \frac{1}{2}k(r-r_o)^2)} dr \nonumber \\ &\sim& 4\pi
r_o^2 \left(\frac{2 \beta}{k}\right)^{1/2} e^{-\beta w_o}
\int_{-\infty}^\infty e^{-t^2}dt \nonumber \\ &=& 4\pi r_o^2  \left
( \frac{2\pi \beta}{k}\right)^{1/2} e^{-\beta w_o}
\label{eq:ncparaasymp}
\end{eqnarray}
which is the usual leading order term in the expansion of a Laplace
integral~\cite{bender}. Therefore, the experimental signature of a
non-central-well is a Langmuir constant which behaves at low
temperatures like
\begin{equation}
C(\beta) \sim C_o e^{m\beta} \beta^{1/2} \mbox{ as } \beta \rightarrow
\infty.
\label{eq:ncpara}
\end{equation}
Comparing (\ref{eq:ncparaasymp}) and (\ref{eq:ncpara}), we can identify
the well depth $w_o=-m$, consistent with the general arguments above,
but it is impossible to determine independently the location of $r_o$ and the
curvature $k$ of the minimum. Instead, any $r_o$ and $k$ satisfying
$4\pi r_o^2 \sqrt{2\pi/k} = C_o$ would exactly reproduce the same
large-$\beta$ asymptotics of the Langmuir curve (as would a completely
different central-well solution described in
section~\ref{sec:deviations}).  This degeneracy of non-central-well
solutions revealed in the low temperature asymptotics is actually
characteristic of all non-central-well solutions, as explained below.

In the central-well case, $r_o=0$, the asymptotics must be carried out
more carefully because the leading term derived in
(\ref{eq:ncparaasymp}) vanishes:
\begin{eqnarray}
C(\beta) &\sim & 4\pi \beta \int_0^\epsilon e^{-\beta(w_o +
\frac{1}{2}k r^2)} r^2 dr \nonumber \\
&\sim& 4\pi \left(\frac{2}{k^3\beta}\right)^{1/2} e^{-\beta w_o}
\int_0^\infty t^{1/2} e^{-t} dt \nonumber \\
&=& \left(\frac{2\pi}{k}\right)^{3/2}\frac{ e^{-\beta
w_o}}{\beta^{1/2}}.
\end{eqnarray}
The experimental signature of a parabolic central well in the Langmuir
curve,
\begin{equation}
C(\beta) \sim C_o e^{m\beta} \beta^{-1/2} \mbox{ as } \beta \rightarrow
\infty,
\label{eq:cpara}
\end{equation}
is qualitatively different from (\ref{eq:ncpara}), which provides an
unambiguous way to separate the two cases using low temperature
measurements.  Moreover, unlike the non-central-well case, the
curvature $k = 2\pi C_o^{-2/3}$ of a parabolic central minimum is
uniquely determined by the low temperature asymptotics of the Langmuir
curve. Consistent with asymptotic results, we shall see in
section~\ref{sec:general} that {\it central-well solutions to the
inverse problem are unique, while non-central-well solutions are not}.

\subsection{Sufficient Conditions for the Existence of Solutions}

The primary difficulty in solving Eq.~(\ref{eq:langmuir}) lies in its
being a nonlinear integral equation of the ``first kind'' for which no
general theory of the existence and uniqueness of solutions
exists~\cite{ww,tricomi}.  In the linear case, however, there is a
special class of first-kind equations which can be solved using
Laplace, Fourier or Mellin transforms, namely integral equations of
the additive or multiplicative convolution
type~\cite{carrier,titchmarsh}
\begin{eqnarray}
\Psi(x) & = & \int_{-\infty}^\infty K(x-y)\psi(y)dy \\
\mbox{ or \ \ \  } \Psi(x) & = & \int_0^\infty K(xy)\psi(y)dy,
\label{eq:convmult}
\end{eqnarray}
respectively, where $\psi(x)$ is the unknown function and $\Psi(x)$ is
given.  Integral equations of the multiplicative form
(\ref{eq:convmult}) often arise in statistical mechanics as explained
in the Introduction.

Although our nonlinear, first-kind equation (\ref{eq:langmuir}) is not
of the convolution type because the unknown function $w(r)$ appears in
the exponent, it does somewhat resemble a Laplace transform. This
connection is more obvious in the alternative formulation (\ref{eq:u})
relating $F(\beta)$ and $u(x)$, which is equivalent to the original
equation (\ref{eq:langmuir}) according to the analysis above. In the
next section, it is shown that physically reasonable solutions exist
whenever $F(\beta)$ has an inverse Laplace transform $f(y)$ which is
positive, nondecreasing and non-constant for $y>0$. (Sufficient
conditions on $F(\beta)$ to ensure these properties of $f(y)$ are
given in Appendix B.)

\section{Analytical Solutions for Arbitrary Langmuir Curves}
\label{sec:general}

\subsection{The Unique Central-Well Solution}

\begin{figure}
\begin{center}
\mbox{
\psfig{file=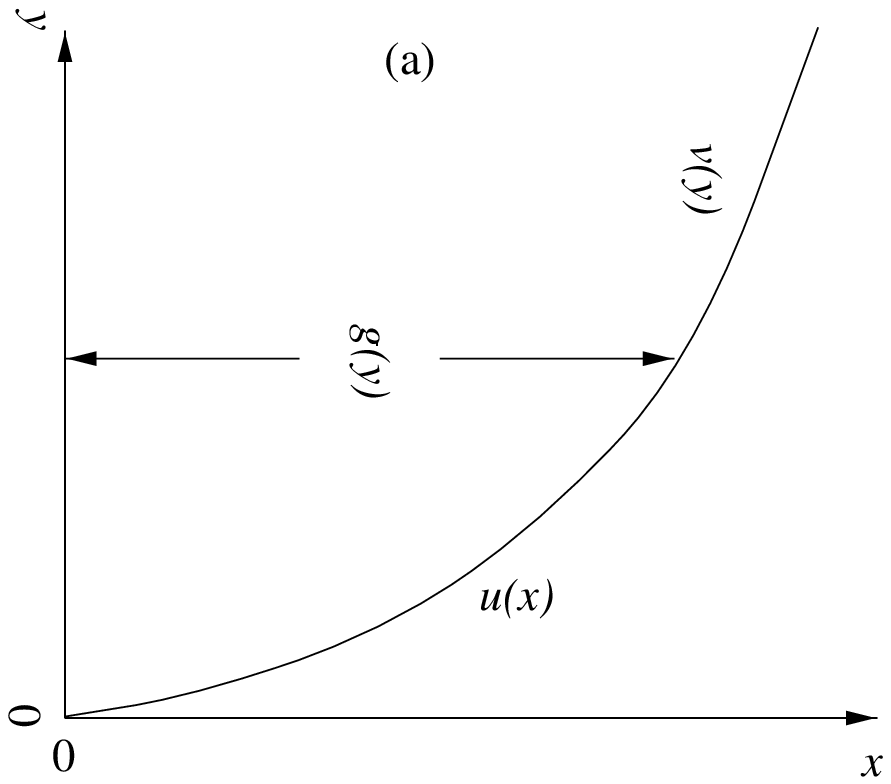,width=3.5in}
} \\
\vspace{0.2in}
\mbox{
\psfig{file=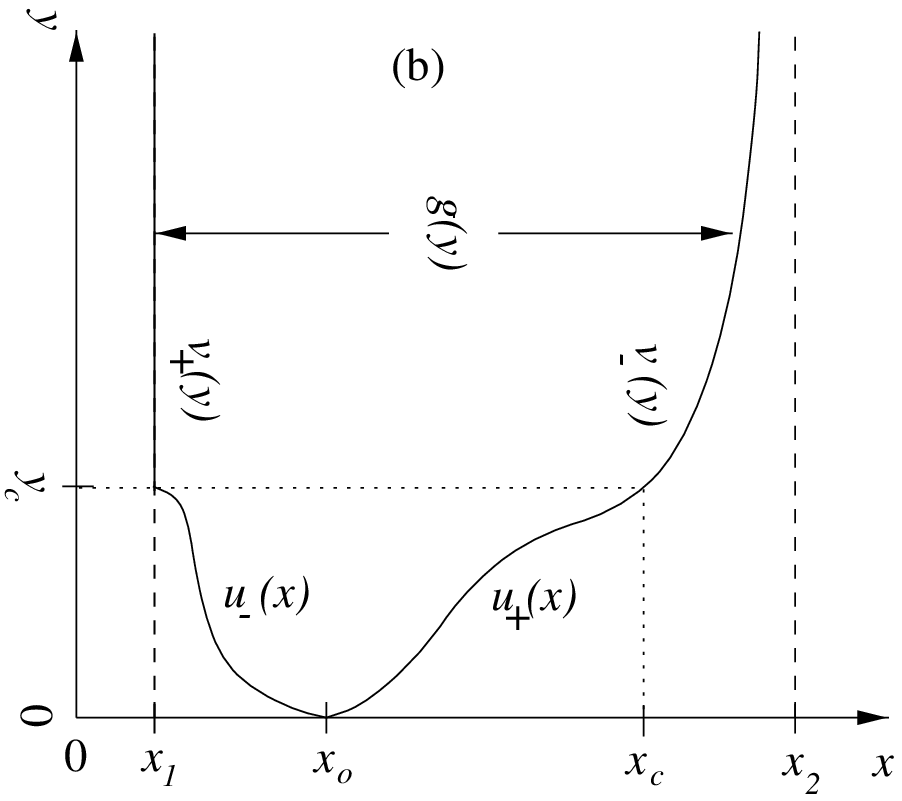,width=3.5in}
}\\
\vspace{0.2in}
\begin{minipage}[h]{5in}
\caption{
(a) Sketch of a central-well solution, where $x$ is the
scaled volume of interaction and $u(x)$ is the spherically averaged
cell potential with inverse $v(y) = u^{-1}(y)$.  (b) Sketch of a
non-central-well solution composed of a non-increasing function $u_-(x)$
and a nondecreasing function $u_+(x)$ joined at a minimum of zero at
$x_o$, along with a possible hard core at $x_1$ and hard wall at
$x_2$.  The two branches $v_-(y)$ and $v_+(y)$ of the multi-valued
inverse cell potential $v(y)$ are also shown, along with other
variables defined in the text.
\label{fig:uv}}
\end{minipage}
\end{center}
\end{figure}

It is tempting to change variables $y=u(x)$ in the integral
(\ref{eq:u}) to reduce it to a Laplace transform, but care must be
taken since $u^{-1}(y)$ may not be single-valued. This leads us to
treat solutions which are monotonic separately from those from those
which are not, an important distinction fore-shadowed by the asymptotic
analysis above.  As a natural first case, we seek differentiable
solutions $u(x)$ which are strictly increasing without bound
($u(\infty)=\infty$) from a central minimum ($u(0)=0$). Such
``central-well solutions'' correspond to cell potentials $w(r)$ which
are strictly increasing from a finite minimum $w(0)=w_o$ at the center
of the cage.  We proceed by considering the inverse cell potential
$v(y) = u^{-1}(y)$ with units of volume as a function of energy, which
is single-valued and strictly increasing with $v(0)=0$, as shown in
Fig.~\ref{fig:uv}(a).  With the substitution $y=u(x)$, Eq.~(\ref{eq:u})
is reduced to Laplace's integral
equation~\cite{titchmarsh} for the unknown function $v^\prime(y)$
\begin{equation}
F(\beta) = \int_0^\infty \frac{e^{-\beta y} dy}{u^\prime(u^{-1}(y))} =
\int_0^\infty e^{-\beta y} v^\prime(y) dy.
\label{eq:F1}
\end{equation}
Upon taking inverse Laplace transforms, we arrive at a differential
equation for $v(y)$
\begin{equation}
v^\prime(y) = f(y)
\label{eq:v1}
\end{equation}
whose unique solution is
\begin{equation}
v(y)= \int_0^y f(y)dy
\end{equation}
using the boundary condition $v(0)=0$.  According to (\ref{eq:v1}),
the continuity of $f(y)$ for $y>0$ (which is not assumed) would the
guarantee differentiability of $v(y)$ for $y>0$, and hence of $u(x)$
for $x>0$.

Equivalently, we can also simplify (\ref{eq:F1}) with an integration
by parts
\begin{equation}
F(\beta) = \beta \int_0^\infty e^{-\beta y} v(y) dy.
\end{equation}
Therefore, the inverse cell potential is given by
\begin{equation}
v(y) = g(y)
\label{eq:vcentral}
\end{equation}
where $g(y)$ is the inverse Laplace transform of the function
\begin{equation}
G(\beta) = \frac{F(\beta)}{\beta} = \frac{C(\beta) e^{\beta w_o}}{
\beta^2}.
\end{equation}
The cell potential $u(x)$ is determined implicitly by the algebraic
equation
\begin{equation}
g(u) = x.
\end{equation}
Returning to the original variables, we have a general expression for
$w(r)$ in the central-well case
\begin{equation}
w(r) = w_o + g^{-1}\left( \frac{4}{3}\pi r^3 \right)
\label{eq:central}.
\end{equation}
{\it This equation uniquely determines the central-well potential that
exactly reproduces any admissible Langmuir curve.}


\subsection{Non-Central-Well Solutions}

The simplest kind of non-central-well solution is the central-well
(\ref{eq:central}) shifted by a ``hard-core'' radius $r_{hc}> 0$
\begin{equation}
w(r) = \left\{ \begin{array}{ll}
\infty & \mbox{ if } 0 \leq r < r_{hc} \\
w_o + g^{-1}\left[ \frac{4}{3}\pi (r^3-r_{hc}^3) \right]
	& \mbox{ if } r \geq r_{hc}
\end{array} \right.
\label{eq:whard}
\end{equation}
which exemplifies a peculiar general property of our integral
equation: An arbitrary hard core can be added to any solution. Note
that, if $u(x)$ is any solution of the rescaled equation (\ref{eq:u}),
then so is
\begin{equation}
\tilde{u}(x) = \left\{ \begin{array}{ll}
\infty & \mbox{ if } 0 \leq x < x_{hc} \\
u(x-x_{hc}) & \mbox{ if } x \geq x_{hc}
\end{array} \right.
\end{equation}
for any hard-core volume $x_{hc} \geq 0$. The proof is simple:
\begin{equation}
\int_0^\infty e^{-\beta \tilde{u}(x)}dx =
\int_{x_{hc}}^\infty e^{-\beta u(x-x_{hc})}dx =
\int_0^\infty e^{-\beta u(x)}dx = F(\beta).
\end{equation}
Physically, a hard core for the cell potential could represent the
presence of a second guest molecule (in a spherically symmetric model)
in the same clathrate-hydrate
cage. Alternatively, a hard-core could represent a water molecule
(again in a spherically symmetric model) at
the node of several adjacent clathrate cages, in which case the cell
potential actually describes the ``super-cage'' surrounding the
central water molecule.

The arbitrary hard-core just described only hints at the vast
multiplicity of non-monotonic solutions to the integral equation
(\ref{eq:u}), which is a common characteristic of first-kind
equations~\cite{tricomi}. Next, we consider the general case of a
non-central-well, shown in Fig.~\ref{fig:uv}(b), which includes
(\ref{eq:whard}) as a special case. To be precise, we seek continuous
solutions $u(x)$ on an interval $x_1 < x < x_2$ composed of a
non-increasing function $u_-(x)$ and a nondecreasing function $u_-(x)$
which are piecewise differentiable and non-negative.  We also allow
for a possible hard-core in the central region $x<x_1$ as well as a
``hard wall'' beyond the clathrate cage boundary $x>x_2$. The general
form of such a non-central-well solution is
\begin{equation}
u(x) =  \left\{ \begin{array}{ll}
\infty & \mbox{ if } 0 < x < x_1 \\
u_-(x) & \mbox{ if } x_1 < x \leq x_o \\
u_+(x) & \mbox{ if } x_o \leq x < x_2 \\
\infty & \mbox{ if } x < x_2 \\
\end{array} \right.
\label{eq:u2}
\end{equation}
where $u_-(x_o) = u_+(x_o)$.  We do not assume
$u^\prime_+(x_o)=u^\prime_-(x_o)$, which would imply differentiability
at the minimum $u^\prime(x_o)=0$, although we do not rule out this
case either.  Instead, we allow for the mathematical possibility of a
discontinuous first derivative at $x_o$, i.e. a ``cusp'' at the well
position, at least for the moment.

As before, it is convenient to express the solution (\ref{eq:u2}) in
terms of two differentiable functions $v_-(y) = u_-^{-1}(y)$ and
$v_+(y) = u_+^{-1}(y)$ which describe the multi-valued inverse cell
potential. Note that $v_-(\infty)=x_1$, $v_+(\infty)=x_2$ and
$v_-(0)=v_+(0)=x_o$. In terms of the inverse cell potentials, the
integral equation (\ref{eq:u}) takes the form
\begin{eqnarray}
F(\beta) &=& \int_0^\infty e^{-\beta u(x)}dx \nonumber \\
&=& \int_{x_1}^{x_o} e^{-\beta u_-(x)}dx + \int_{x_o}^{x_2}
e^{-\beta u_+(x)}dx \nonumber \\
&=& \int_0^\infty e^{-\beta y} \left[v^\prime_+(y) -
v_-^\prime(y)\right] dy
\label{eq:vv}
\end{eqnarray}
which implies
\begin{equation}
v^\prime_+(y) - v_-^\prime(y) = f(y).
\label{eq:vp}
\end{equation}
In this case, the continuity of $f(y)$ would only guarantee the
differentiability of the difference $v_+(y) - v_-(y)$, but not of the
individual functions $v_+(y)$ and $v_-(y)$. Integrating (\ref{eq:vv})
by parts before taking the inverse transform yields a general
expression for the solution
\begin{equation}
v_+(y) - v_-(y) = g(y)
\label{eq:noncentral}
\end{equation}
where again $g(y)$ is the inverse Laplace transform of
$F(\beta)/\beta$.  Unfortunately, we have two unknown functions and
only one equation, so the set of non-central-well solutions is
infinite.

The scaled Langmuir curve $F(\beta)$ uniquely determines only $v_+(y)
- v_-(y)$, the volume difference as a function of energy between the
two branches of the scaled cell potential $u(x)$, but not the branches
$v_+(y)$ and $v_-(y)$ themselves. An infinite variety of
non-central-well solutions, which exactly reproduce the same Langmuir
curve as the central well solution, can be easily generated by
choosing any non-increasing, non-negative, piecewise differentiable
function $v_-(y)$ such that the function $v_+(y)$ defined by
(\ref{eq:noncentral}) is nondecreasing.  Even the position of the well
$v_-(0)=x_o$ can be chosen arbitrarily.

For example, one such family of solutions with a central ``soft-core''
($x_1=0$) is given by
\begin{equation}
u(x) = \left\{ \begin{array}{ll}
u_-(x) & \mbox{ if } 0 \leq x \leq x_o \\
u_+(x) & \mbox{ if } x \geq x_o
\label{eq:ugensoft}
\end{array} \right.
\end{equation}
where
\begin{equation}
v_-(y) = u^{-1}_-(y) = \left\{ \begin{array}{ll}
x_o - a y^b & \mbox{ if } 0 \leq y \leq y_c \\
0 & \mbox{ if } y \geq y_c
\end{array} \right.
\end{equation}
and
\begin{equation}
v_+(y) = u_+^{-1}(y) = v_-(y) + g(y),
\end{equation}
for any $a, b >0$ and $x_o \geq 0$. (In the limit $a \rightarrow 0$,
we recover the unique central-well solution.) Note that $y_c = u(0) =
(x_o/a)^{1/b}$ is the height of the central maximum of the potential.
These solutions, all derived from a single Langmuir curve, exist
whenever $g(y)$ increases quickly enough that $v_+(y)$ is
nondecreasing, which is guaranteed if $g^\prime(y) \geq a b y^{b-1}$
for $0 < y <y_c$.

Another family of non-central-well solutions with a soft-core can be
constructed with the choice
\begin{equation}
v_-(y) = \left\{ \begin{array}{ll}
x_o - \lambda g(y) & \mbox{ if } 0 \leq y \leq y_c \\
0 & \mbox{ if } y \geq y_c
\end{array} \right.
\end{equation}
for any $0 < \lambda < 1$ and $x_o \geq 0$, where $y_c =
g^{-1}(x_o/\lambda)$.  In this case, the cell potential is easily
expressed in terms of $g^{-1}(x)$ as
\begin{equation}
u(x) = \left\{ \begin{array}{ll}
g^{-1}\left(\frac{x_o-x}{\lambda}\right) & \mbox{ if } 0\leq x \leq x_o\\
g^{-1}\left(\frac{x_o-x}{1-\lambda}\right) & \mbox{ if } x_o\leq x \leq x_c\\
g^{-1}(x) & \mbox{ if } x \geq x_c
\end{array}\right.
\label{eq:uncsoft}
\end{equation}
where $x_c = v_+(y_c) = x_o/\lambda$. This class of solutions exists
whenever $g(y)$ is nondecreasing (or $f(y) \geq 0$). If $y_c =
\infty$, then there is a hard core $u(0)=\infty$. Otherwise, if there
is a soft core $u(0)=y_c<\infty$, then there is typically a cusp
(discontinuous derivative) at $x_c$, as explained below.

As demonstrated by the preceding examples, it is simple to generate an
enormous variety of non-central-well solutions, with an arbitrarily
shaped soft or hard core, and an arbitrary position of the minimum.
In spite of the multiplicity of non-central-well solutions, however,
our analysis of the inverse problem at least determines $v_+(y) -
v_-(y)$ uniquely from any experimental Langmuir curve. {\it This
important analytical constraint is not satisfied by empirical fitting
procedures.}

\subsection{Soft Cores and Outer Cusps}

Non-central-well solutions with a soft-core satisfy $v_-(y)=0$ for
$y\geq y_c>0$, as in the examples above. In such cases, $v_+(y) = g(y)$
for $y\geq y_c$ regardless of whether or not there is an outer hard
wall, which implies that $u(x) = g^{-1}(x)$ for $x \geq x_c$, where
$x_c = g^{-1}(y_c)$. If $f(y)$ is continuous for $y>0$, then, unless
$u_-(x)$ has an ``inverted cusp'' at the origin ($v_-^\prime(y_c)=0$
and $u_-^\prime(0^+)=-\infty$), any non-central-well solution $u(x)$
with a soft-core must have a cusp at $x=x_c$, as in the examples
above. This ``outer cusp'' in $u(x)$ could only be avoided if $f(y)$
itself has a cusp at $y_c$ which would allow $v_+(y)$ to be
continuous.  However, an inverted cusp in $u(x)$ at the origin does
not necessarily imply an cusp in $w(r)$ at the origin due the
transformation $x = \frac{4}{3}\pi r^3$. For example, if $v_-(y) \sim
(y_c - y)^{3/2}$ as $ y \rightarrow y_c$, or $u(x) \sim y_c - x^{2/3}$
as $x \rightarrow 0$, then $w(r)$ would have a physically reasonable,
parabolic soft core $w(r) \sim w_o + y_c - (4\pi/3)^{2/3} r^2$ as
$r\rightarrow 0$. Nevertheless, even in such cases, if $f(y)$ were
continuous for all $y>0$, then both $u(x)$ and $w(r)$ would have
unphysical second-derivative discontinuities at $x=x_c$ related to the
soft core. In general, a continuously differentiable, non-central-well
solution with a central soft core could only arise if $f(y)$ were
discontinuous at some $y_c>0$, and such discontinuities are generally
not present.

\subsection{Cusps at a Non-Central Minimum}

As mentioned above, the behavior of the cell potential near its
minimum (whether central or not) is determined by the behavior of the
Langmuir curve at low temperature, or equivalently, at large inverse
temperature, $\beta=1/T$. The Laplace transform formalism makes this
connection transparent and mathematically rigorous. The asymptotic
behavior of $G(\beta)$ as $\mbox{Re}\beta\rightarrow\infty$ is related
to the asymptotics of the inverse transform $g(y)$ as $y \rightarrow
0$, which in turn governs the local shape of the energy minimum
through Eq.~(\ref{eq:vcentral}) for a central well or
Eq.~(\ref{eq:noncentral}) for a non-central well.  The leading order
asymptotics has already been computed above for parabolic minima, but
the general solutions above show how various non-local properties of
the potential are related to finite temperature features of the
Langmuir curve. Here, we comment on a subtle difference in
differentiability between central and non-central-wells, related to
the small-$y$ behavior of $f(y)$.

For typical sets of experimental data, including the van't Hoff form
(\ref{eq:Cexpt}), the prefactor $F(\beta)$ has a bounded inverse
Laplace transform in the neighborhood of the origin
\begin{equation}
\lim_{y \rightarrow 0} f(y) = f(0) <\infty.
\label{eq:bound}
\end{equation}
This generally implies the existence of a cusp at a non-central
minimum of $u(x)$, which is signified by a nonzero right and/or left
derivative.  When $u(x)$ is differentiable at its minimum, it
satisfies $u_-^\prime(x_o^-)=u_+^\prime(x_o^+)=0$. In the central-well
case $x_o=0$, the existence of a cusp in $u(x)$ follows from
(\ref{eq:v1})
\begin{equation}
u^\prime(0^+) = 1/v^\prime(0^+) = 1/f(0) > 0,
\end{equation}
but this does not imply a cusp in the unscaled potential $w(r)$ as
long as $f(0)>0$ because in that case
\begin{equation}
w(r) - w_o = u(4\pi r^3/3) \sim (4\pi/3f(0)) r^3 \mbox{ as } r
\rightarrow 0^+.
\end{equation}
In the non-central-well case, however, the bounded inverse transform
(\ref{eq:bound}) implies a cusp at the minimum because, $v_+^\prime(0)
- v_-^\prime(0) = f(0) < \infty$ from (\ref{eq:noncentral}) along with
$v_+^\prime(0)\geq 0$ and $v_-^\prime(0)\leq 0$ implies that
$v_+^\prime(0)<\infty$ and/or $v_-^\prime(0)>-\infty$ which in turn
implies $u_+^\prime(0)>0$ and/or $u_-^\prime(0)<0$. Unlike the
central-well case, however, a cusp in $u(x)$ at the non-central minimum
$x_o>0$ implies a cusp at the corresponding non-central minimum of
$w(r)$.  Therefore, we conclude that whenever (\ref{eq:bound}) holds,
the only physically reasonable solution is the central-well solution
(\ref{eq:central}).

\subsection{Asymptotics at High Energy and Temperature}

The high energy behavior of the cell potential is related to (but not
completely determined by) the high temperature asymptotics of the
Langmuir hydrate constant, through the function $g(y)$.  For example,
the cell potential would have a hard wall at $x_2 < \infty$, if and
only if $g(y)$ were unbounded
\begin{equation}
\lim_{y\rightarrow\infty} g(y) = \infty.
\label{eq:ginf}
\end{equation}
Since the empirical modeling of Langmuir curves using Kihara
potentials assumes an outer hard-core, Eq.~(\ref{eq:ginf}) could be
used to test the suitability of using the Kihara potential form,
although experimental data is often not available at sufficiently high
temperatures to make a fully adequate comparison
(see below). Whenever (\ref{eq:ginf}) holds, the
non-central-well solutions $u(x)$ are also universally asymptotic to
the central-well solution
\begin{equation}
u(x) \sim g^{-1}(x)
\label{eq:largex}
\end{equation}
at large volumes $x\rightarrow\infty$. This follows from
(\ref{eq:noncentral}) and the fact that $v_-(y)$ is bounded, which
implies $v_+(y) \sim g(y)$. The exact inversions performed in
section~\ref{sec:deviations} provide further insight into the
relationship between small $\beta$ asymptotics of the Langmuir
constant and high energy behavior of the cell potential.

\section{Langmuir Curves with van't Hoff Temperature Dependence}
\label{sec:vanthoff}

Experimental Langmuir hydrate-constant curves $C(\beta)$ are well fit
by an ideal van't Hoff temperature dependence (\ref{eq:Cexpt}),
demonstrated by straight lines on Arrhenius log-linear plots
\begin{equation}
\log C = m \beta + \log C_o
\label{eq:VH}
\end{equation}
as shown in Figs.~\ref{fig:Cexpt-linear} and \ref{fig:Cexpt} for
ethane ($C_o=4.733\times 10^{-7}$ atm$^{-1}$, $m=9.4236$ kcal/mol) and
cyclopropane ($C_o=1.9041\times 10^{-7}$ atm$^{-1}$, $m=10.5939$
kcal/mol) clathrate-hydrates~\cite{sparks_thesis}. This data is
analyzed carefully in section~\ref{sec:expt}, where alternative
functional forms are considered.  In the ideal van't Hoff case, we have
$F(\beta) = C_o/\beta$ and $G(\beta) = C_o/\beta^2$. The inverse
Laplace transforms of these functions are simply $f(y) = C_o H(y)$ and
$g(y) = C_o y H(y)$, respectively, where $H(y)$ is the Heaviside step
function.

\begin{figure}
\begin{center}
\mbox{
\psfig{file=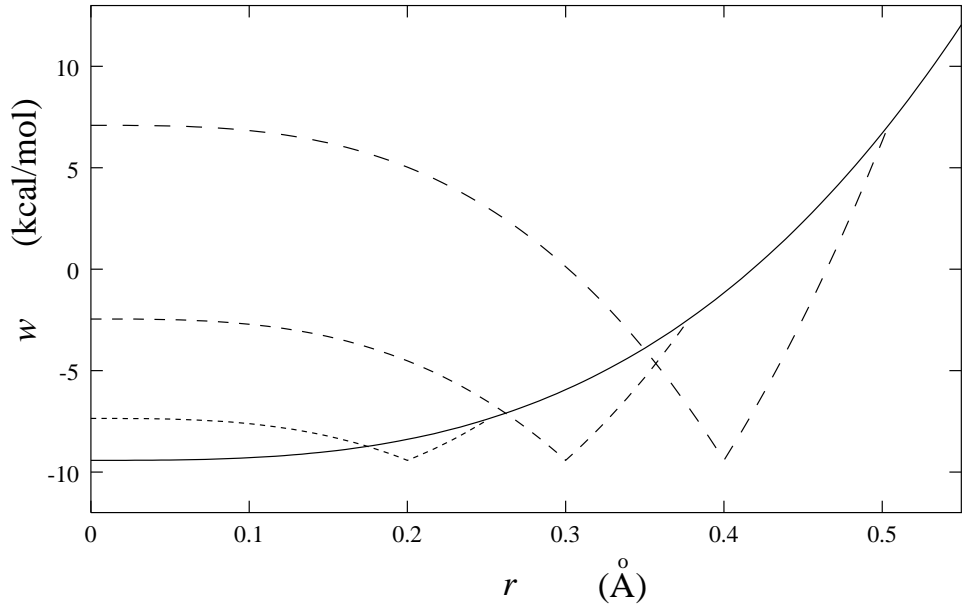,width=5in}
}\\
\vspace{0.2in}
\begin{minipage}[h]{5in}
\caption{
Analytical cell potentials for the ethane clathrate-hydrate
which exactly reproduce the experimental data in
Fig.~\ref{fig:Cexpt}(a). The unique central-well solution
(\protect\ref{eq:wcentral}) is indicated by a solid line, while a
family of non-central-well solutions with soft cores
(\protect\ref{eq:wsoft}) is also shown as dashed lines with cusp-like
minima at $r_o = 0.2,0.3,0.4$ \AA. Each of these solutions also has a
cusp at $r=2^{1/3} r_o$, where the energy is the same as the central
maximum, and beyond this distance joins the central-well solution.
\label{fig:w-ethane}
}
\end{minipage}
\end{center}
\end{figure}

We begin by discussing the unique central-well solution, which is
illustrated by the solid line in Fig.~\ref{fig:w-ethane} for the case
of ethane. The central-well solution is linear in volume $u(x) = g(x)
= C_o y H(y)$, and cubic in radius
\begin{equation}
w(r) = \frac{4\pi r^3}{3 C_o} - m.
\label{eq:wcentral}
\end{equation}
A curious feature of this exact solution is that it has a vanishing
``elastic constant'', $w^{\prime\prime}(0)=0$, a somewhat unphysical
property which we address again in section~\ref{sec:expt}.

The simple form of (\ref{eq:wcentral}) makes it very appealing as a
means of interpreting experimental data with van't Hoff temperature
dependence. We have already noted that the slope of a van't Hoff (Fig. 2) plot
of the Langmuir constant is equal to the well depth $m=-w_o$, but now
we see that the $y$-intercept $\log C_o$ is related to the well-size,
e.g. measured by the volume of negative energy $m C_o$. This volume
corresponds to a spherical radius of
\begin{equation}
r_s = \left( \frac{3m C_o}{4\pi}\right)^{1/3}
\label{eq:rs}
\end{equation}
which is $0.4180$ \AA \ for ethane and $0.3208$ \AA \ for
cyclopropane. Since the van der Waals radius of ethane is less than
that of cyclopropane, it makes physical sense that $r_{s}^{\rm ethane}
> r_{s}^{\rm cyclopropropane}$.  Moreover, these volumes fall within
the ranges determined from two different experimental modeling
approaches: Using the radius of the water cage from x-ray scattering
experiments~\cite{sloan}, Lennard-Jones potentials from gas viscosity
data give 0.79 \AA\ for ethane and 0.61 \AA\ for
cyclopropane~\cite{RPP}, while computations with van der Waals radii
give 0.18 \AA\ for ethane and 0.03 \AA\ for
cyclopropane~\cite{sloan}.

There are infinitely many non-central-well solutions reproducing van't
Hoff temperature dependence, but each of them has unphysical cusps
(discontinuous derivatives). There will always be a cusp at the
minimum of the potential, since $f(y)$ satisfies the general condition
(\ref{eq:bound}).  For example, the central-well solution can be
shifted by an arbitrary hard-core radius $r_o \geq 0$
\begin{equation}
w(r) = \left\{ \begin{array}{ll} \infty & \mbox{ if } 0 \leq r < r_o
\\ \frac{4\pi (r^3 - r_o^3)}{3 C_o} - m & \mbox{ if } r\geq r_o
\end{array} \right.
\end{equation}
In the case of a soft core, there must be a second cusp in the outer
branch of the potential at the same energy as the inner core due to
the continuity of $f(y)$, as explained above. This is illustrated by
the following piecewise cubic family of soft-core solutions of the
general form (\ref{eq:uncsoft}):
\begin{equation}
w(r) = \left\{ \begin{array}{ll} \frac{8\pi|r_o^3 - r^3|}{3 C_o} - m &
\mbox{ if } 0 \leq r \leq 2^{1/3} r_o \\ \frac{4\pi r^3}{3 C_o} - m &
\mbox{ if } r_o \geq 2^{1/3} r_o
\label{eq:wsoft}
\end{array}\right.
\end{equation}
which are shown in Fig.~\ref{fig:w-ethane} in the case of ethane guest
molecules. An infinite variety of other piecewise differentiable
solutions exactly reproducing van't Hoff dependence of the Langmuir
curve could easily be generated, as described above, but each would
have unphysical cusps.

Previous studies involving {\it ad hoc} fitting of Kihara potentials
have reported non-central-wells~\cite{sloan}, but these empirical fits
may be only approximating various exact, cusp-like, non-central-well
solutions, such as those described above. Moreover, given that the
central-well solution (\ref{eq:wcentral}) can perfectly reproduce the
experimental data, it is clear that the results obtained by fitting
Kihara potentials to Langmuir curves are simply artifacts of the {\it
ad hoc} functional form, without any physical significance. Kihara
fits also assume a hard wall at the boundary of the clathrate cage (by
construction), whereas all of the exact analytical solutions (both
central and non-central-wells) have the asymptotic dependence
\begin{equation}
w(r) \sim \frac{4\pi r^3}{3 C_o}
\end{equation}
as $r \rightarrow \infty$ according to (\ref{eq:largex}). Any
deviation from the cubic shape at large radii, such as a hard wall,
would be indicated by a deviation from van't Hoff behavior at high
temperatures, but such data would be difficult to attain in
experiments (see below).

The preceding analysis shows that the only physical information
contained in a Langmuir curve with van't Hoff temperature
dependence is the depth $w_o$ and the effective radius $r_s$ of the
spherically averaged cell potential, which takes the unique form
(\ref{eq:wcentral}) in the central-well case. In hindsight, the simple
two-parameter form of the potential is not surprising since a van't
Hoff dependence is described by only two parameters, $m$ and $C_o$. It
is clearly inappropriate to fit more complicated {\it ad hoc}
functional forms, such as Eq.~(\ref{eq:w-Kihara}) derived from the
Kihara potential, since they contain extraneous fitting parameters and
do not reproduce the precise shape of any exact solution.


\section{Analysis of Possible Deviations from van't Hoff Behavior}
\label{sec:deviations}

\subsection{Dimensionless Formulation}

The general analysis above makes it possible to predict analytically
the significance of possible deviations from van't Hoff temperature
dependence, which could be present in the experimental data (see
below). We have already discussed the experimental signatures of
various low and high energy features of the cell potential in the
asymptotics of the Langmuir curve. In this section, we derive exact
solutions for Langmuir curves of the form (\ref{eq:Fdef}) where
$F(\beta)$ is a rational function. Such cases correspond to
logarithmic corrections of linear behavior on a van't Hoff plot of the
Langmuir curve, which are small enough over the accessible temperature
range to be of experimental relevance, in spite of the dominant van't
Hoff behavior seen in the data.

Fitting to the dominant van't Hoff behavior (\ref{eq:VH}) introduces
natural scales for energy, $m$, and pressure, $C_o^{-1}$, so it is
convenient and enlightening to introduce dimensionless variables.
With the definitions
\begin{equation}
\tilde{\beta} = m\beta, \ \  \tilde{C}(\tilde{\beta}) =
C(\tilde{\beta}/m)/C_o, \ \mbox{ and } \
\tilde{F}(\tilde{\beta}) = F(\tilde{\beta}/m)/mC_o,
\end{equation}
the Langmuir curve can be expressed in the dimensionless form
\begin{equation}
\tilde{C}(\tilde{\beta}) = \tilde{\beta}\tilde{F}(\tilde{\beta})
e^{\tilde{\beta}}.
\end{equation}
For consistency with these definitions, the other energy-related
functions in the analysis are nondimensionalized as follows
\begin{equation}
\tilde{G}(\tilde{\beta}) = G(\tilde{\beta}/m)/m^2C_o, \ \
\tilde{y} = y/m, \ \ \tilde{f}(\tilde{y}) = f(m\tilde{y})/C_o, \ \
\tilde{g}(\tilde{y}) = g(m\tilde{y})/mC_o,
\end{equation}
where $\tilde{f}(\tilde{y})$ and $\tilde{g}(\tilde{y})$ are the
inverse Laplace transforms of $\tilde{F}(\tilde{\beta})$ and
$\tilde{G}(\tilde{\beta})$, respectively.  The natural scales for
energy and pressure also imply natural scales for volume, $mC_o$, and
distance, $r_s$, as described in the previous section, which motivates
the following definitions of the dimensionless cell potential versus
volume
\begin{equation}
\tilde{x} = x/mC_o, \ \ \tilde{u}(\tilde{x}) = u(mC_o \tilde{x})/m
\end{equation}
and radius
\begin{equation}
\tilde{r} = r/r_s, \ \ \tilde{w}(\tilde{r}) = w(r_s \tilde{r})/m.
\end{equation}
Note that $\tilde{x} = \tilde{r}^3$. With these definitions, the
central-well solution takes the simple form,
\begin{equation}
\tilde{u}(\tilde{x}) = \tilde{g}^{-1}(\tilde{x})
\end{equation}
in terms of the dimensionless volume, or
\begin{equation}
\tilde{w}(\tilde{r}) = -1 + \tilde{g}^{-1}(\tilde{r}^3)
\end{equation}
in terms of the dimensionless radius.  We now consider various
prefactors $\tilde{F}(\tilde{\beta})$ which encode valuable
information about the energy landscape in various regions of the
clathrate cage.

\subsection{The Interior of the Clathrate Cage}

\subsubsection{Power-Law Prefactors}

\begin{figure}
\begin{center}
\mbox{
\psfig{file=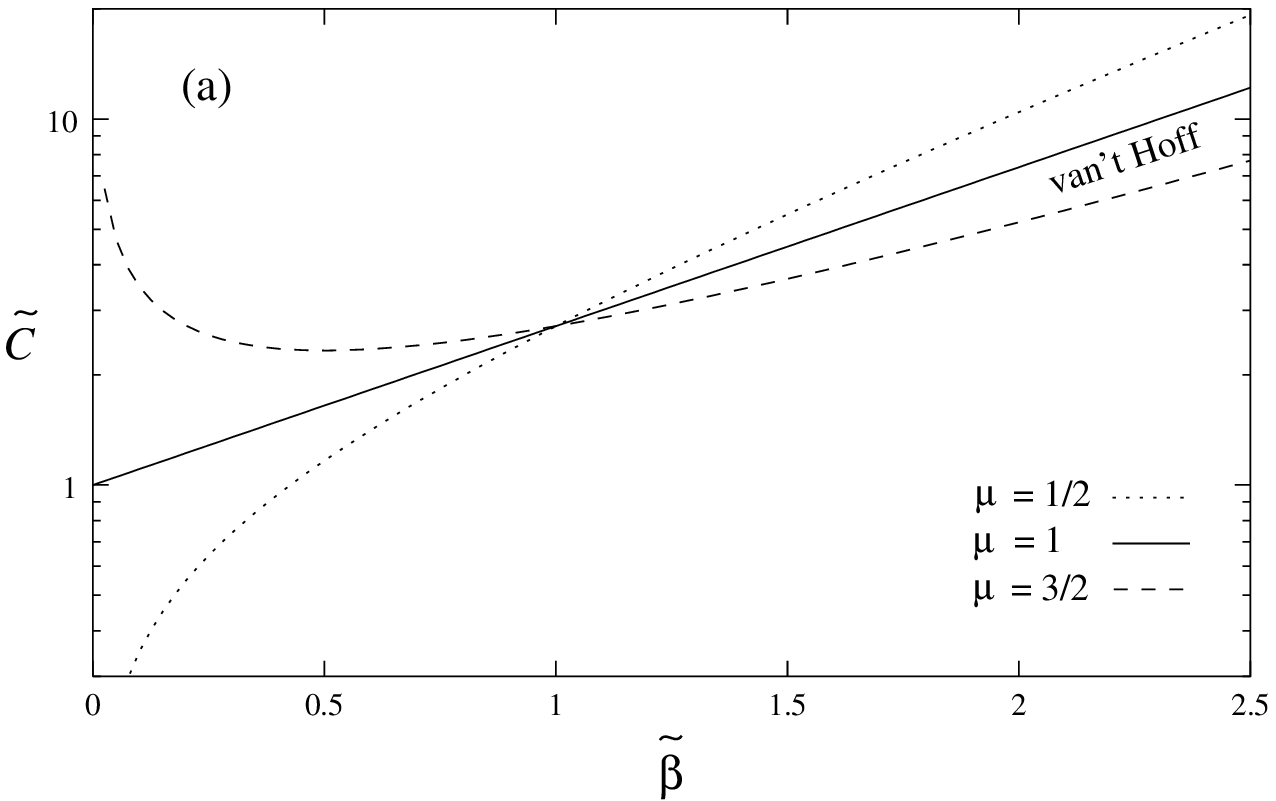,height=2.4in}
} \\
\vspace{0.2in}
\mbox{
\psfig{file=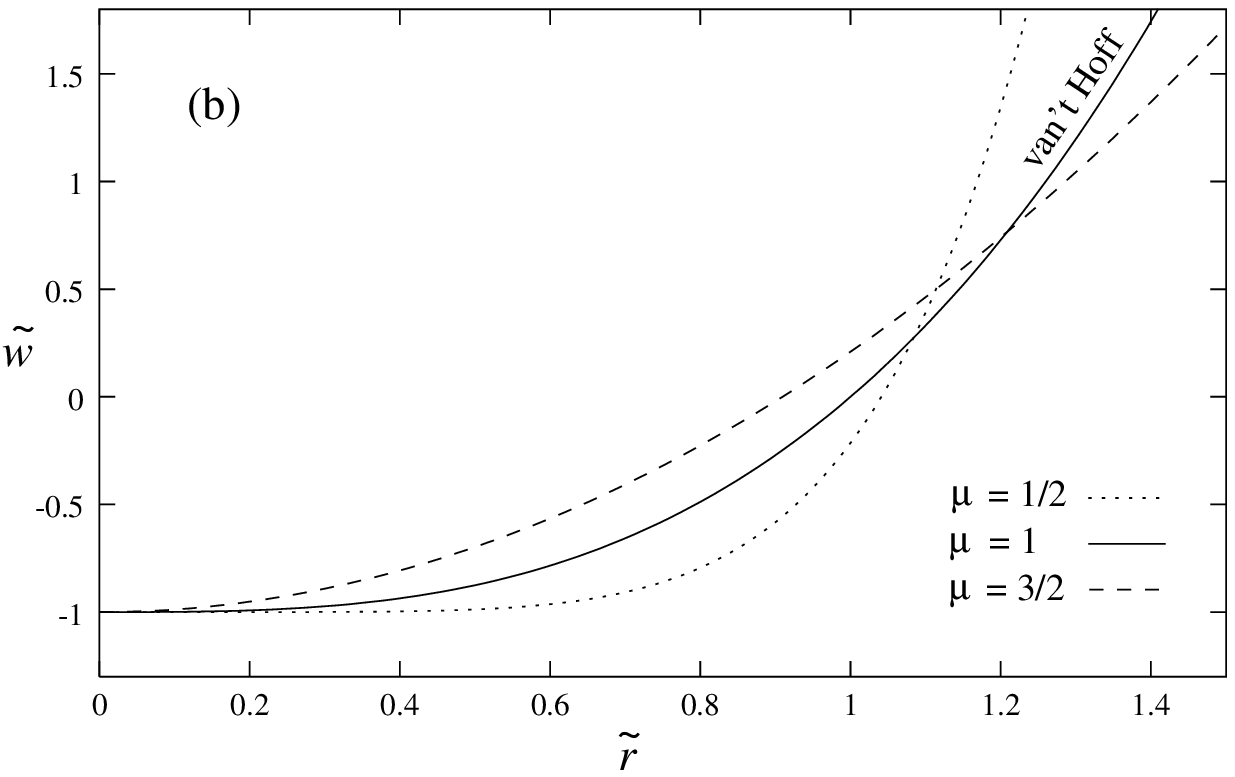,height=2.4in}
} \\
\vspace{0.2in}
\mbox{
\psfig{file=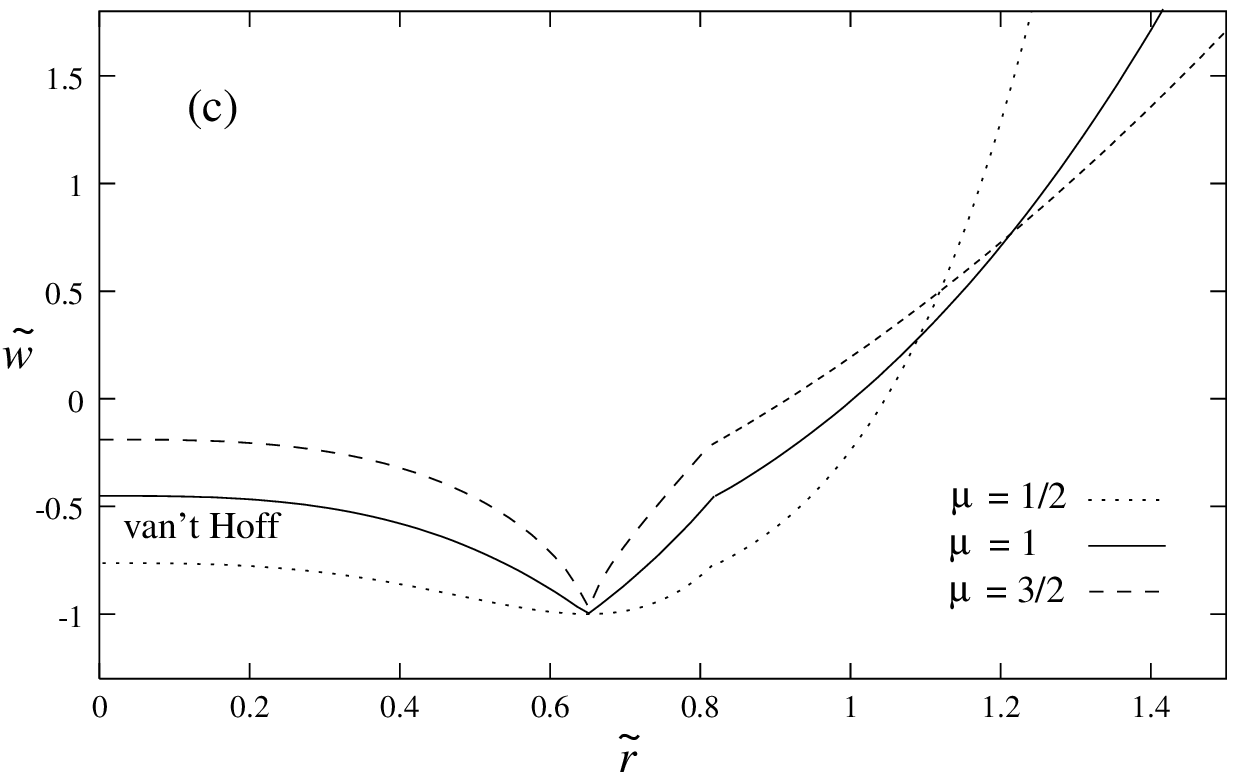,height=2.4in}
}\\
\vspace{0.2in}
\begin{minipage}[h]{5in}
\caption{
Exact inversion of Langmuir curves with power-law corrections
to van't Hoff behavior in terms of dimensionless variables, as in
Eq.~(\protect\ref{eq:c-powerlaw}). (a) Plots of $\tilde{C} = C/C_o$
versus $\tilde{\beta} = m/kT$ for the cases $\mu=1/2, 1, 3/2$. (b) The
corresponding (unique) central-well potentials plotted as $\tilde{w} =
w/m$ versus $\tilde{r} = r/(3mC_o/4\pi)^{1/3}$. (c) Examples of
soft-core non-central-well solutions of the form
(\ref{eq:wnc-powerlaw}) with an arbitrarily chosen minimum at
$\tilde{r}=0.65$, which all have cusps at
$\tilde{r}=2^{1/3}(0.65)\approx 0.819$.
\label{fig:c-powerlaw}
}
\end{minipage}
\end{center}
\end{figure}

The simplest possible correction to van't Hoff behavior involves a
power-law prefactor
\begin{equation}
\tilde{F}(\tilde{\beta}) = \tilde{\beta}^{-\mu} \ \ \mbox{ for any }
\mu>0,
\end{equation}
which corresponds to a logarithmic correction on a van't Hoff
plot of the Langmuir constant
\begin{equation}
\log \tilde{C} = \tilde{\beta} + (1-\mu) \log(\tilde{\beta})
\label{eq:c-powerlaw}
\end{equation}
as shown in Fig.~\ref{fig:c-powerlaw}(a). In this case, we have
\begin{equation}
\tilde{f}(\tilde{y}) = \tilde{y}^{\mu-1} H(\tilde{y})/\Gamma(\mu)
\ \mbox { and } \
\tilde{g}(\tilde{y}) = \tilde{y}^{\mu} H(\tilde{y})/\Gamma(\mu+1),
\end{equation}
where $\Gamma(z)$ is the gamma function. In general, power-law
prefactors at low temperatures signify an energy minimum with a simple
polynomial shape.

\subsubsection{ The Central-Well Solution }

The unique central-well solution is also a simple power law
\begin{equation}
\tilde{u}(\tilde{x}) = \left[\Gamma(\mu+1) \tilde{x}\right]^{1/\mu}
\end{equation}
or equivalently
\begin{equation}
\tilde{w}(\tilde{r}) = -1 + \Gamma(\mu+1)^{1/\mu} \tilde{r}^{3/\mu}.
\label{eq:wcmu}
\end{equation}
The cubic van't Hoff behavior is recovered in the case $\mu=1$, as is
the (asymptotic) parabolic behavior from (\ref{eq:wpara}) and
(\ref{eq:cpara}) in the case $\mu=3/2$.  Because
$\tilde{w}(\tilde{r})+1 \propto \tilde{r}^{3/\mu}$, a power-law
correction to van't Hoff behavior with a positive exponent ($\mu<1$)
corresponds one which is ``wider'' than a cubic, while a negative
exponent ($\mu>1$) corresponds to a potential which is ``more narrow''
than a cubic, as shown in Fig.~\ref{fig:c-powerlaw}(b).  On physical
grounds, the smooth polynomial behavior described by (\ref{eq:wcmu})
is always to be expected near the minimum energy of the cell
potential. Therefore, the power-law correction to van't Hoff behavior
(\ref{eq:c-powerlaw}) has greatest relevance for low temperature
measurements in the range $\tilde{\beta} \gg 1$, from which it
determines interatomic forces in the interior of the clathrate cage at
low energies $|\tilde{w}(\tilde{r})| \ll 1$.

\subsubsection{ Non-Central-Well Solutions }

As described above, there are infinitely many non-central-well
solutions.  One family of solutions of the form (\ref{eq:uncsoft})
with $\lambda=1/2$ is given by
\begin{equation}
\tilde{w}(\tilde{r}) + 1 = \left\{ \begin{array}{ll}
\left[ 2 \Gamma(\mu+1) |\tilde{r}^3 - \tilde{r}_o^3|\right]^{1/\mu}
   & \mbox{ if } 0\leq \tilde{r} \leq 2^{1/3}\tilde{r}_o \\
\left[ \Gamma(\mu+1) \tilde{r}^3 \right]^{1/\mu}
   & \mbox{ if } \tilde{r} \geq 2^{1/3}\tilde{r}_o
\end{array} \right.
\label{eq:wnc-powerlaw}
\end{equation}
where $\tilde{r}_o = r_o/r_s$ is arbitrary, as shown in
Fig.~\ref{fig:c-powerlaw}(c) for the case $\tilde{r}_o=0.65$. These
solutions are unphysical since they all have cusps at
$\tilde{r} = 2^{1/3}\tilde{r}_o$ near the outer wall of the
cage. However, they can still have reasonable behavior near the
minimum at $\tilde{r}_o$ for certain values of $\mu$, which could have
experimental relevance for low temperature measurements. Near the
minimum, the exact solutions (\ref{eq:wnc-powerlaw}) have the
asymptotic form
\begin{equation}
\tilde{w}(\tilde{r}) \sim -1 + \left[ 6 \Gamma(\mu+1) \tilde{r}_o^2
|\tilde{r} - \tilde{r}_o| \right]^{1/\mu} \ \ \mbox{ as } \tilde{r}
\rightarrow \tilde{r}_o,
\end{equation}
which is cusp-like for $\mu>1/2$, but differentiable for $0<\mu\leq
1/2$. For example, the non-central-well has a parabolic shape in the
case $\mu=1/2$, which agrees with the asymptotic analysis in
Eqs.~(\ref{eq:wpara})--(\ref{eq:ncpara}) when the units are restored, and
it has a cubic shape when $\mu=1/3$.  On the other hand, in the
central-well case $\mu=3/2$ and $\mu=1$ correspond to parabolic and
cubic minima, respectively. Therefore, this example nicely illustrates
the difference between the low-energy asymptotics of central and
non-central-wells described above in section~\ref{sec:math}, which
would be useful in interpreting any experimental Langmuir constant
data showing deviations from van't Hoff behavior.

\subsection{The Outer Wall of the Clathrate Cage}

\subsubsection{ Rational Function Prefactors }

The behavior of the Langmuir curve in the high temperature region
$\tilde{\beta} = O(1)$ is directly linked to properties of the outer
wall of the clathrate cage, described by the cell potential at high
energies $\tilde{w}(\tilde{r})+1 = O(1)$. Although this region of the
Langmuir curve does not appear to be accessible in experiments (see
below), in this section we derive exact solutions possessing different
kinds of outer walls, whose faint signature might someday be observed
in experiments at moderate temperatures. In order to isolate possible
effects of the outer wall, we consider Langmuir curves which are
exactly asymptotic to the usual van't Hoff behavior at low temperatures
with small logarithmic corrections (on a van't Hoff plot) at moderate
temperatures. These constraints suggest choosing rational functions
for $\tilde{F}(\tilde{\beta})$ such that $\tilde{F}(\tilde{\beta})
\sim 1/\tilde{\beta}$ as $\beta \rightarrow \infty$.

\subsubsection{Central Wells with Hard Walls}

\begin{figure}
\begin{center}
\mbox{
\psfig{file=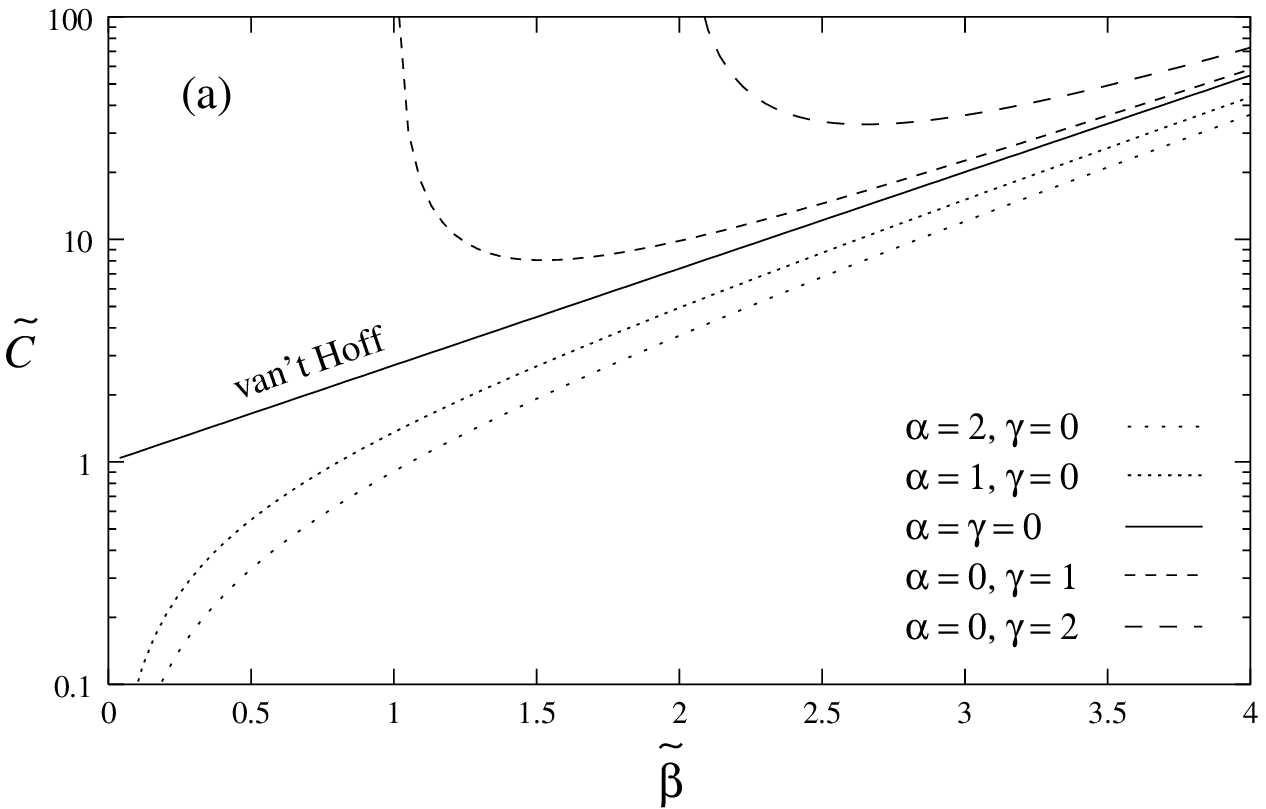,width=5in}
} \\
\vspace{0.2in}
\mbox{
\psfig{file=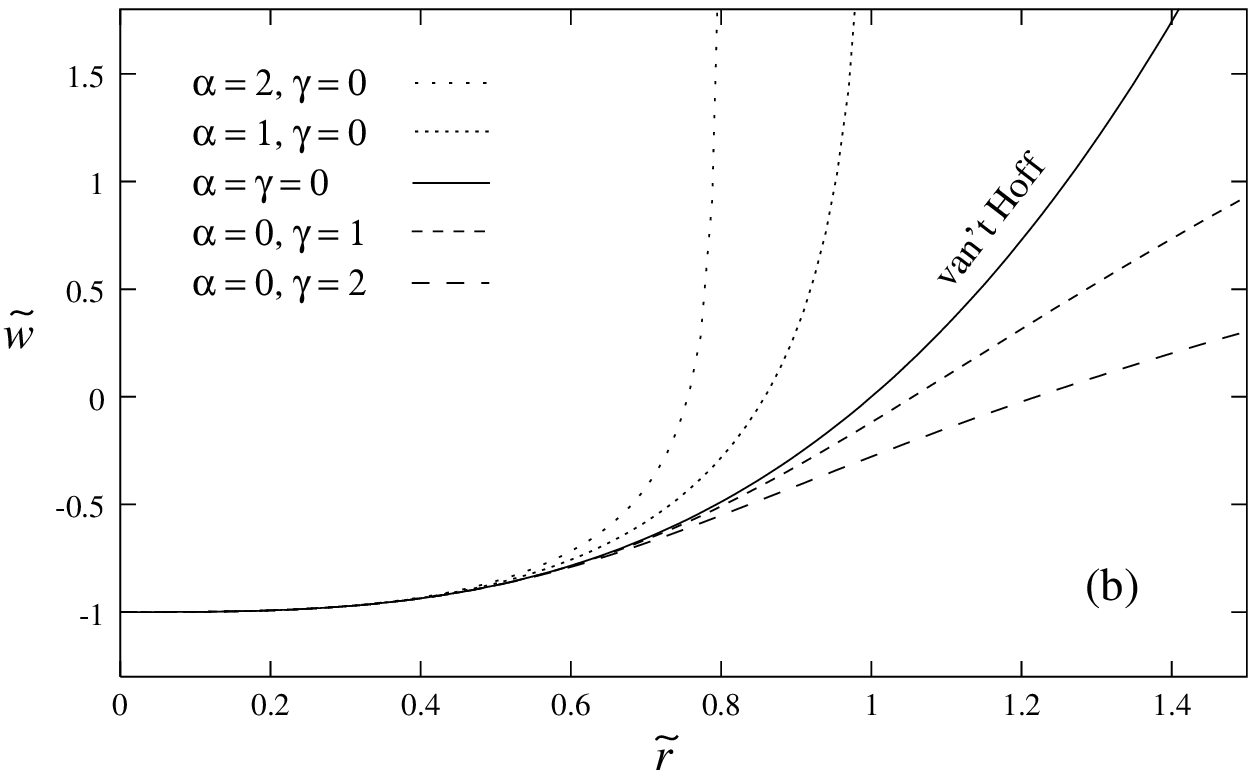,width=5in}
}\\
\vspace{0.2in}
\begin{minipage}[h]{5in}
\caption{
Exact inversion of Langmuir curves of the form $\tilde{C} =
e^{\tilde{\beta}}/(1 + \alpha/\tilde{\beta} -
(\gamma/\tilde{\beta})^2)$ in terms of the dimensionless variables
defined in Fig.~\ref{fig:c-powerlaw}. (a) Langmuir curves in this
class of functions have anomalous high temperature (small
$\tilde{\beta}$) behavior but are asymptotic to the van't Hoff curve
($\alpha=\gamma=0$). (b) The corresponding central-well solutions
depart from the cubic van't Hoff curve at large radius and energy,
indicating different properties at the boundary of the clathrate cage:
``hard walls'', if $\alpha>0$ and $\gamma=0$, or ``soft walls'', if
$\alpha=0$ and $\gamma>0$.
\label{fig:cag}
}
\end{minipage}
\end{center}
\end{figure}

We begin by considering a ``shifted power-law'' prefactor
\begin{equation}
\tilde{F}(\tilde{\beta}) = 1/(\tilde{\beta} + \alpha), \ \
\mbox{ for any } \alpha>0
\end{equation}
which corresponds to a shifted logarithmic deviation from van't Hoff
behavior,
\begin{equation}
\log \tilde{C} = \tilde{\beta} - \log(1 + \alpha/\tilde{\beta}).
\label{eq:calpha}
\end{equation}
As shown in Fig.~\ref{fig:cag}(a), this suppresses the Langmuir
constant at high temperatures, which intuitively should be connected
with an enhancement of the strength of the outer wall compared to the
cubic van't Hoff solution.  Taking inverse Laplace transforms
we have
\begin{equation}
\tilde{f}(\tilde{y}) = e^{-\alpha \tilde{y}} H(\tilde{y})
\ \mbox { and } \
\tilde{g}(\tilde{y}) = (1 - e^{-\alpha \tilde{y}}) H(\tilde{y})/\alpha,
\end{equation}
and indeed, since $\tilde{g}(\tilde{y})$ is bounded, all solutions
must have a hard wall regardless of whether or not the well is
central, as described above. For example, the unique central-well
solution is
\begin{equation}
\tilde{w}(\tilde{r}) = -1 - \log(1 - \alpha \tilde{r}^3)/\alpha \ \
\mbox{ for } 0 \leq \tilde{r} < \alpha^{-1/3}
\label{eq:wchardwall}
\end{equation}
which has an outer hard wall at $\tilde{r} = \alpha^{-1/3}$, as shown
in Fig.~\ref{fig:cag}(b). The solution is also asymptotic to the cubic
van't Hoff solution at small radii $\tilde{r} \ll
\alpha^{-1/3}$. Therefore, in the limit $\alpha \rightarrow 0$, the
radius of the outer hard wall diverges, and the solution reduces to
the cubic shape as the deviation from van't Hoff behavior is moved to
increasingly large temperatures. Since empirical fitting with Kihara
potential forms arbitrarily assumes an outer hard wall, this example
provides analytical insight into the nature of the approximation at
moderate to high temperatures, where the Langmuir constant should be
suppressed according to (\ref{eq:calpha}).

\subsubsection{ Central Wells with Soft Walls }

Next we consider the opposite case of a Langmuir constant which is
enhanced at high temperatures compared to van't Hoff behavior, which
intuitively should indicate the presence of a ``soft wall'', rising
much less steeply than a cubic function. An convenient
choice is
\begin{equation}
\tilde{F}(\tilde{\beta}) = \tilde{\beta}/(\tilde{\beta}^2 -
\gamma^2), \ \  \mbox{ for any } \gamma>0.
\end{equation}
which is analytic except for poles at $\beta = \pm \gamma$ on the real
axis.  Although this function diverges at $\beta=\gamma$ due to the
overly soft outer wall, the corresponding Langmuir curve
\begin{equation}
\log \tilde{C} = \tilde{\beta} - \log\left[ 1 -
(\gamma/\tilde{\beta})^2\right]
\label{eq:cgamma}
\end{equation}
shown in Fig.~\ref{fig:cag}(a) could have experimental relevance at
moderate temperatures $\tilde{\beta} \gg \gamma$, if $\gamma$ were
sufficiently small. In this case, we have
\begin{equation}
\tilde{f}(\tilde{y}) = \cosh(\gamma \tilde{y}) H(\tilde{y})
\ \mbox{ and } \
\tilde{g}(\tilde{y}) = \sinh(\gamma \tilde{y}) H(\tilde{y})/\gamma
\end{equation}
which yields the central-well solution
\begin{equation}
\tilde{w}(\tilde{r}) = -1 + \sinh^{-1}(\gamma \tilde{r}^3)/\gamma.
\label{eq:wcsoftwall}
\end{equation}
As shown in Fig.~\ref{fig:cag}(b), this function follows the van't Hoff
cubic at small radii $\tilde{r} \ll \gamma^{-1/3}$ but ``softens'' to
a logarithmic dependence for large radii $\tilde{r} \gg
\gamma^{-1/3}$.

\section{Interpretation of Experimental Data}
\label{sec:expt}

\begin{figure}
\begin{center}
\mbox{
\psfig{file=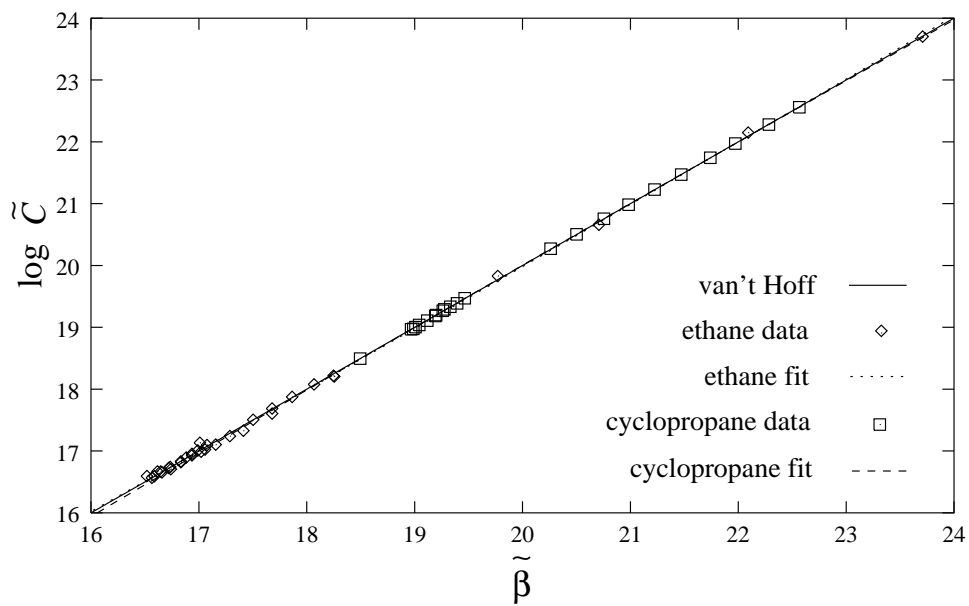,width=5in}
}\\
\vspace{0.2in}
\begin{minipage}[h]{5in}
\caption{ 
Collapse of the experimental Langmuir curves for ethane and
cyclopropane hydrates from Fig.~\protect\ref{fig:Cexpt} plotted in
terms of the dimensionless variables $\tilde{C} = C/C_o$ versus
$\tilde{\beta} = m/kT$, where $C_o$ and $m$ are obtained by fitting
each data set to $\log C = m/kT + \log C_o$. Ideal van't Hoff behavior
$\tilde{C} = \exp(\tilde{\beta})$ is shown as a solid line. Fits
including power law corrections as in Eq.~(\protect\ref{eq:C-fit}) are
also shown as the dotted and dashed lines (which are very close to the
van't Hoff line).\label{fig:C-fit}
}
\end{minipage}
\end{center}
\end{figure}

We begin by fitting Langmuir curves, computed from
experimental phase equilibria data, an equation of state, and
reference thermodynamic properties~\cite{sparks_thesis} for
ethane and cyclopropane clathrate-hydrates to the van't Hoff equation
\begin{equation}
\log C = m \beta + b
\label{eq:VHfit}
\end{equation}
using least-squares linear regression. This leads to rather accurate
results, as indicated by the small uncertainties in the parameters
displayed in Table~\ref{tab:fit} (63\% confidence intervals
corresponding to much less than one percent error). The high quality
of the regression of $\log C$ on $\beta$ is further indicated by
correlation coefficients very close to unity, $0.99650$ and $0.99998$
for the ethane and cyclopropane data, respectively.  Using the fitted
values for $m$ and $C_o = e^b$, the data for the two
clathrate-hydrates can be combined into a single plot in terms of the
dimensionless variables $\tilde{C}$ and $\tilde{\beta}$, as shown in
in Fig.~\ref{fig:C-fit}, which further demonstrates the common linear
dependence.

Converting the experimental data to dimensionless variables also
reveals that the measurements correspond to extremely ``low
temperatures''. This is indicated by large values of $\tilde{\beta} =
m/kT$ in the range of 16 to 24, which imply that $kT$ is less than $6\%$
of the well depth $m$. As such, physical intuition tells us that the
experiments can probe the cell potential only very close to its
minimum. This intuition is firmly supported by the asymptotic analysis
above, which (converted to dimensionless variables) links the
asymptotics of the Langmuir constant for $\tilde{\beta} \gg 1$ to that
of the cell potential for $|\tilde{r} - \tilde{r}_o| \ll 1$.  In this
light, it is clear that any features of the cell potential other than
the local shape of its minimum, which are determined by empirical
fitting, e.g. using Eq.~(\ref{eq:w-Kihara}) based on the Kihara
potential, are simply artifacts of an {\it ad hoc} functional form,
devoid of any physical significance.

Since the shape of the potential very close to its minimum should
always be well approximated by a polynomial (the leading term in its
Taylor expansion), the analysis above implies that only simple
power-law prefactors to van't Hoff behavior should be considered in
fitting low temperature data. Therefore, we refit the experimental
data, allowing for a logarithmic correction,
\begin{equation}
\log C = m \beta + b + \nu \log (\beta)
\label{eq:C-fit}
\end{equation}
as in Eq.~(\ref{eq:c-powerlaw}).  The results are shown in
Table~\ref{tab:fit}, and the best-fit functions are displayed in
dimensionless form in Fig.~\ref{fig:C-fit}.  In the case of ethane,
the best-fit value of $\mu=1-\nu$ corresponds to a roughly linear
central-well solution $\tilde{w} \propto \tilde{r}^{0.9}$ or a
cusp-like non-central-well solution $\tilde{w} \propto |\tilde{t} -
\tilde{r}_o|^{0.3}$. Although these solutions are not physically
reasonable, perhaps the qualitative increase in $\mu$ compared with
ideal van't Hoff behavior ($\mu=1$) is indicative of a parabolic
central well ($\mu=3/2$). In the case of cyclopropane, we have $\mu =
-1.4 \pm 0.9$, which violates the general condition $\mu\geq 0$ needed
for the existence of solutions to the inverse problem. If this fit
were deemed reliable, then the basic postulate of vdWP theory,
Eq.~(\ref{eq:Phi}), would be directly contradicted, with or without
the spherical cell approximation (see the Appendix B). It is perhaps
more likely that the trend of decreasing $\mu < 1$ could indicate a
non-central parabolic minimum in the spherically averaged cell
potential ($\mu=1/2$).

Although it appears there may be systematic deviations from ideal van't
Hoff behavior in the experimental data for ethane and cyclopropane,
$\nu \neq 0$ or $\mu \neq 1$, the results are statistically
ambiguous. For both types of guest molecules, adding the third degree
of freedom $\nu$ substantially degrades the accuracy of the two
linear parameters $m$ and $b$, with errors increased by
several hundred percent. Moreover, the uncertainty in $\nu$ is
comparable to its best-fit value.  Therefore, it seems that we cannot trust
the results with $\nu \neq 0$, and, by the principle of Occam's
razor, we are left with the more parsimonious two-parameter fit to
van't Hoff behavior, which after all is quite good, and its associated
simple cubic, central-well solution.

On the other hand, there are different two-parameter fits, motivated
by the inversion theory, which can describe the experimental data
equally well, but which are somewhat more appealing than the cubic
solution in that they possess a non-vanishing elastic constant (second
spatial derivative of the energy). For
example, the fits can be done using (\ref{eq:C-fit}) with the
parameter $\nu$ fixed at either $1/2$ or $-1/2$, corresponding to
either a non-central or central, parabolic minimum, respectively. The
results shown in Table~\ref{tab:fit} reveal that these physically
significant changes in the functional form have little effect on the
van't Hoff parameters $m$ and $b = \log C_o$.

The difficulty with the experimental data as a starting point for
inversion is its limited range in $\tilde{\beta}$ of roughly one
decade, which makes it nearly impossible to detect corrections
proportional to $\log \beta$ related to different polynomial shapes of
the minimum. It would be very useful to extend the range of the data,
using the analytical predictions to interpret the results. In general,
it is notoriously difficult to determine power-law prefactors
multiplying a dominant exponential dependence, but at least the
present analysis provides important guidance regarding the appropriate
fitting functions, which could not be obtained by {\it ad hoc}
numerical fitting. Moreover, the clear physical meaning of the
dominant van't Hoff parameters elucidated by the analysis also makes
them much more suitable to describe experimental data than the
artificial Kihara potential parameters.

\section{Summary}
\label{sec:concl}

In this article, we have shown that spherically averaged
intermolecular potentials can be determined analytically from the
temperature dependence of Langmuir constants. Starting from the
statistical theory of van der Waals and Platteeuw, the method has been
developed for the case of clathrate-hydrates which contain a single
type of guest molecule occupying a single type of cage. Finally, the
method has been applied to experimental data for ethane and
cyclopropane clathrate-hydrates. Various conclusions of the analysis
are summarized below.

\vskip 12pt
\noindent {\bf General Theoretical Conclusions}

\begin{itemize}

\item Physically reasonable intermolecular potentials (which are
piecewise continuous and bounded below) exist only if the Langmuir
curve has a dominant exponential (van't Hoff) dependence at low
temperatures, $\lim_{\beta \rightarrow \infty} \log C/\beta = m$, with
a prefactor $F(\beta) = C(\beta)e^{-m\beta}/\beta$ which is smooth and
non-increasing.

\item The slope $m$ of an experimental ``van't Hoff plot'' of $\log C$
versus inverse temperature $\beta$ is precisely equal to the well
depth, i.e. (minus) the minimum of the potential. This is true not
only for the spherically averaged cell potential, $\min w(r) = -m$,
but also for the exact multi-dimensional potential, $\min \Phi(\vec{r})
= -m$.

\item For any physically reasonable Langmuir curve, the unique
central-well potential can be determined from Eq.~(\ref{eq:central}).

\item There also exist infinitely many non-central-well solutions of
the form (\ref{eq:u2}), constrained only to satisfy
Eq.~(\ref{eq:noncentral}). Several classes of such solutions with a
central ``soft core'' (a finite maximum at the center of the cage) are
described explicitly in Eqs.~(\ref{eq:ugensoft})--(\ref{eq:uncsoft}).

\item Each one of the multitude of non-central-well solutions with a
soft-core typically possesses unphysical cusps (slope
discontinuities), while the unique central-well solution is a
well-behaved analytic function.

\item For ideal van't Hoff temperature dependence, $C(\beta) = C_o
e^{m\beta}$, the central-well solution is a simple cubic given by
Eq.~(\ref{eq:wcentral}). The attractive region of the potential has
depth $m$, volume $mC_o$, and radius $r_s=(3mC_o/4\pi)^{1/3}$. Each
non-central-well solution for van't Hoff dependence has two unphysical
cusps, one at the minimum.

\item The experimental signature of a parabolic, non-central-well is a
Langmuir curve that behaves like $C(\beta) \sim C_o e^{m\beta}
\beta^{1/2}$ at low temperatures ($\beta \rightarrow
\infty$), while a parabolic central well corresponds to $C(\beta) \sim
C_o e^{m\beta} \beta^{-1/2}$.

\item If there is a pure power-law prefactor multiplying van't Hoff
behavior $C(\beta) = C_o (m\beta)^{1-\mu} e^{m\beta}$ with $\mu>0$,
the central-well solution is also a power-law (\ref{eq:wcmu}). For
certain values of the prefactor exponent $0 < \mu \leq 1/2$, there are
also non-central-well solutions with differentiable minima such as
(\ref{eq:wnc-powerlaw}), although such solutions still possess
cusps at higher energies.

\item Rational function prefactors multiplying van't Hoff behavior,
such as (\ref{eq:calpha}) or (\ref{eq:cgamma}), are associated with
non-cubic behavior at the outer wall of the cage, such as a ``hard
wall'' (\ref{eq:wchardwall}) or a ``soft wall'' (\ref{eq:wcsoftwall}),
respectively.

\end{itemize}

\vskip 12pt

\noindent {\bf Conclusions for Clathrate-Hydrates}

\begin{itemize}

\item Since Langmuir constants must increase with temperature, $m>0$,
on the basis of general thermodynamical arguments, the intermolecular
potential must be attractive (with a region of negative energy).

\item The depth $w_o$ and radius $r_s$ of the attractive region of the
cell potential can be estimated directly from experimental data using
the simple formulae $w_o = -m$ and $r_s=(3mC_o/4\pi)^{1/3}$ without
any numerical fitting.
The resulting values for ethane and cyclopropane hydrates are
consistent with typical estimates obtained by other means.

\item The experimental Langmuir constant data for ethane and
cyclopropane clathrate-hydrates is very well fit by an ideal van't
Hoff dependence, which corresponds to a cubic central well, 
\[
w(r) = \frac{4\pi r^3}{3 C_o} - m
\]
as given by Eq.~(\ref{eq:wcentral}).  However, the data is also
equally consistent with a central parabolic well,
\[
w(r) = \frac{\pi r^2}{C_o^{2/3}} - m,
\]
or various non-central (spherically averaged) parabolic wells. The
range of temperatures is insufficient to distinguish between these
cases.

\item Experimental data tends to be taken at very ``low''
temperatures, $kT \ll m$, which means that only the region of the
potential very close to the minimum $|r - r_o| \ll r_s$ is
probed. Therefore, only simple polynomial functions are to be
expected, and fitting to more complicated functional forms, such as
the Kihara potential, has little physical significance.

\item In practical applications to clathrate-hydrates, the full power
of our analysis could be exploited by measuring Langmuir hydrate
constants over a broader range of temperatures than has previously
been done.

\item The availability of the inversion method obviates the need for
empirical fitting procedures~\cite{sloan,pp}, at least for
single-component hydrates in which guest molecules occupy only one
type of cage.  Moreover, the method also allows a systematic analysis
of empirical functional forms, such as the Kihara potential, which
cannot be expected to have much predictive power beyond the data sets
used in parameter fitting.

\item The general method of ``exact inversion'' developed here could
also be applied to other multi-phase chemical systems, including
guest-molecule adsorption at solid surfaces or in bulk liquids.

\end{itemize}

\section*{Acknowledgments}
We would like to thank Z. Cao for help with the experimental figures,
J. W. Tester for comments on the manuscript, and H. Cheng for useful
discussions.  This work was supported in part by the Idaho National
Engineering and Environmental Laboratory.

\section*{ Appendix A: Inversion of Cohesive Energy Curves for Solids}

The basic idea of obtaining interatomic potentials by ``exact
inversion'' has also recently been pursued in solid state physics
(albeit based on a very different mathematical formalism having
nothing to do with statistical mechanics). The inversion approach was
pioneered by Carlsson, Gelatt and Ehrenreich in 1980 in the case of
pair potentials for crystalline metals~\cite{cge,carlsson}.  These
authors had the following insight: Assuming that the total
(zero-temperature) cohesive energy $E(x)$ of a crystal with nearest
neighbor distance $x$ can be expressed as a lattice sum over all pairs
of atoms $(i,j)$
\begin{equation}
E(x) = \sum_{ij} \phi(x s_{ij})
\label{eq:pair}
\end{equation}
where $s_{ij}$ are normalized atomic separation distances, then a
unique pair potential $\phi(r)$ can be derived which exactly
reproduces the cohesive energy curve $E(x)$. (Lattice sums also appear
in some clathrate-hydrates models~\cite{sparks92}, but to our
knowledge they have never been used as the basis for an inversion
procedure.)

The mathematical theory for the inversion of cohesive energy curves
has been developed considerably in recent years and applied to wide
variety of solids~\cite{chen,chenren,chen94,bk1,bk2}. The extension of
the inversion formalism to semiconductors has required solving a
nonlinear generalization of Eq.~(\ref{eq:pair}) representing many-body
angle-dependent interactions~\cite{bk1,bk2}:
\begin{equation}
F(x) = \sum_{ijk} g(x s_{ij}) g(x s_{ik}) h(\theta_{ijk})
\end{equation}
where $F(x)$ is the many-body energy, $h(\theta_{ijk})$ is the energy
of the angle between two covalent bonds $\vec{r}_{ij} = x
\vec{s}_{ij}$ and $\vec{r}_{ik}= x \vec{s}_{ik}$, and $g(r)$ is a
radial function which sets the range of the interaction. In general,
$\phi(r)$, $g(r)$, and $h(\theta)$ can be systematically obtained from
a set of multiple cohesive energy curves for the same
material~\cite{bk1,bk2,edip}. The angular interaction can also be
obtained directly from cohesive energy curves for non-isotopic
strains~\cite{bazant-thesis}.

\section*{ Appendix B: Mathematical Theorems}

The first theorem provides necessary conditions on the Langmuir
$C(\beta)$ so that the cell potential $w(r)$ is bounded below and
continuous. It also interprets the slope of a van't Hoff plot of
the Langmuir curve in the low temperature limit as the well depth,
under very general conditions. As pointed out in the main text, it is
convenient to view the inverse temperature $\beta$ as a complex
variable.

\begin{theorem}
\label{thm:F}
Let $w(r)$ be real and continuous (except at possibly a finite number
of discontinuities) for $r \geq 0$ with a finite minimum, $w(r) \geq
w_o = w(r_o) > -\infty$ for some $r_o \geq 0$, and suppose that the
integral
\begin{equation}
C(\beta) = 4\pi\beta \int_0^\infty e^{-\beta w(r)}r^2 dr
\label{eq:lang2}
\end{equation}
converges for some $\beta=c$ on the real axis. Then
\begin{equation}
C(\beta) = \beta F(\beta) e^{-w_o\beta}
\label{eq:CF}
\end{equation}
where the complex function $F(\beta)$ is
\begin{itemize}
\item[(i)] real, positive and non-increasing on
the real axis for $\beta > c$ and
\item[(ii)] analytic in the half plane
$\mbox{Re} \beta> c$.
\end{itemize}
If, in addition, the set $S_\epsilon = \{ r\geq 0 | w_o < w(r) < w_o +
\epsilon \}$ has nonzero measure for some $\epsilon = \epsilon_o > 0$,
then $F(\beta)$ is strictly decreasing on the positive real axis (for
$\beta>c$). Moreover, if $S_\epsilon$ has finite, nonzero measure for
every $0 < \epsilon < \epsilon_o$, then
\begin{equation}
\lim_{\beta \rightarrow \infty} \log C(\beta)/\beta = -w_o
\label{eq:asymp}
\end{equation}
where the limit is taken on the real axis.
\end{theorem}

\noindent {\sc Proof:} Define a shifted cell potential versus volume,
$u\left(\frac{4\pi}{3}r^3\right) = w(r) - w_o$. Substituting $u(x)$
for $w(r)$ reduces Eq.~(\ref{eq:lang2}) to Eq.~(\ref{eq:CF}), where
\begin{equation}
F(\beta) = \int_0^\infty e^{-\beta u(x)} dx.
\label{eq:Fintegral}
\end{equation}
Since $u(x) \geq 0$ is real, the function $F(\beta)$ is real and
positive for all real $\beta$ for which the integral
converges. Moreover, for any complex $\beta$ and $\beta^\prime$ with
$\mbox{Re} \beta > \mbox{Re}\beta^\prime > c$, we have the bound
\begin{equation}
|F(\beta)| \leq \int_0^\infty e^{-\mbox{\small Re}\beta \cdot u(x)} dx
\leq \int_0^\infty e^{-\mbox{\small Re}\beta^\prime \cdot u(x)} dx
\leq F(c) < \infty
\end{equation}
which establishes that the defining integral (\ref{eq:Fintegral})
converges in the right half plane $\mbox{Re}\beta \geq c$ and is
non-increasing on the real axis, thus completing the proof of (i).

Next let $w(r)$ be larger than its minimum value (but finite), $w(r_o)
< w(r) < \infty$, on a set $S_\infty$ of nonzero measure, so that $0 <
u(x) < \infty$ for the corresponding set of volumes.  Then for every
$\beta > \beta^\prime > c$ on the real axis we have
\begin{equation}
\int_{S_\infty} e^{-\beta u(x)} dx < \int_{S_\infty} e^{-\beta^\prime
u(x)} dx.
\label{eq:Sinf}
\end{equation}
On the complement ${S_\infty^c} = (0,\infty)\setminus S_\infty$,
either $u(x) = 0$ or $u(x)=\infty$, which implies
\begin{equation}
\int_{S_\infty^c} e^{-\beta u(x)} dx =
\int_{S_\infty^c} e^{-\beta^\prime u(x)} dx.
\label{eq:Sinfc}
\end{equation}
>From Eqs.~(\ref{eq:Sinf})--(\ref{eq:Sinfc}) we conclude that
$F(\beta)$ is strictly decreasing on the real axis.

Next we establish the low-temperature limit (\ref{eq:asymp}).  Given
$0 < \epsilon < \epsilon_o$, we have the following lower bound for any
$\beta > c$ on the real axis:
\begin{equation}
e^{\epsilon \beta} F(\beta) \ = \ \int_0^\infty
e^{-\beta[u(x)-\epsilon]}dx \ \geq \ e^{\epsilon\beta/2}
\int_{S_{\epsilon/2}} dx + \int_{S_{\epsilon/2}^c}
e^{-\beta[u(x)-\epsilon]}dx \  \geq \  e^{\epsilon\beta/2}
\int_{S_{\epsilon/2}} dx.
\end{equation}
Combining this with the upper bound, $F(\beta) \leq F(c) < \infty$, we
obtain
\begin{equation}
e^{-\epsilon/2} M_\epsilon \leq F(\beta) \leq F(c)
\label{eq:Fsqueeze}
\end{equation}
where $M_\epsilon = \int_{S_{\epsilon/2}} dx$ is a finite, nonzero
constant (because $S_{\epsilon/2}$ is assumed to have finite, nonzero
measure). Substituting Eq.~(\ref{eq:CF}) in Eq.~(\ref{eq:Fsqueeze}), we
arrive at
\begin{equation}
\log \beta + \log M_\epsilon - \epsilon \beta/2 - \beta w_o
\ \leq \ \log C(\beta) \ \leq \ \log \beta + \log F(c) - \beta w_o
\end{equation}
which yields
\begin{equation}
-w_o - \epsilon/2 \ \leq \ \lim_{\beta \rightarrow 0} \log C(\beta)/\beta
\ \leq \ -w_o.
\end{equation}
The desired result is obtained in the limit $\epsilon \rightarrow 0$.

Finally, we establish the analyticity of $F(\beta)$ in the open half
plane $\mbox{Re}\beta > c$ by showing that its derivative exists and
is given explicitly by
\begin{equation}
F^\prime(\beta) = -\int_0^\infty e^{-\beta u(x)} u(x)dx.
\label{eq:dF}
\end{equation}
This requires justifying the passing a derivative inside the integral
(\ref{eq:u}), which we have just shown to converge for $\mbox{Re}\beta
\geq c$.  Using a classical theorem of analysis~\cite{ww}, it suffices
to show that the integral in (\ref{eq:dF}) converges uniformly for
$\mbox{Re}\beta > c+\epsilon$ for every $\epsilon > 0$ because the
integrand is a continuous function of $\beta$ and $x$.  (The
possibility of a finite number of discontinuities in $u(x)$ is easily
handled by expressing (\ref{eq:dF}) as finite sum of integrals with
continuous integrands.)  It is a simple calculus exercise to show that
$t e^{-t} < 1/e$, and hence
\begin{equation}
t e^{-(c+\epsilon)t} \leq \frac{e^{-ct}}{e\epsilon}
\end{equation}
for all real $t \geq 0$. This allows us to derive a bound on the
``tail'' of the integral (\ref{eq:dF}):
\begin{equation}
|\int_X^\infty e^{-\beta u(x)} u(x)dx |
\leq \int_X^\infty
e^{-\mbox{{\small Re}}\beta \cdot u(x)} u(x)dx
\leq \int_X^\infty e^{-(c+\epsilon) u(x)} u(x)dx
< \frac{1}{e\epsilon} \int_X^\infty e^{-c u(x)} dx
\end{equation}
which is independent of $\beta$. This uniform bound vanishes in the
limit $X \rightarrow \infty$ because it is proportional to the tail of
the convergent integral defining $F(\beta)$, which completes the
proof.
$\Box$

\vskip 12pt

The proof of Theorem 1 does not depend in any way on the
dimensionality of the integral and thus can be trivially extended to
the general multi-dimensional case of vdWP theory without the
spherical cell approximation.

\begin{theorem}
Let $\Phi(r,\theta,\phi,\alpha,\xi,\gamma) \geq
\Phi(r_o,\theta_o,\phi_o,\alpha_o,\xi_o,\gamma_o) = w_o$ be real and
continuous, and suppose that the integral
\begin{equation}
C(\beta) = \frac{\beta}{8\pi^{2}}\int_V e^{-
\beta \Phi(r,\theta,\phi,\alpha,\xi,\gamma)} r^{2} sin \theta sin \xi dr
d\theta d\phi d\alpha d\xi d\gamma
\label{eq:vdWP2}
\end{equation}
converges for some $\beta = c$ (real). Then all the conclusions of
Theorem 1 hold.
\end{theorem}

\vskip 12pt

The six-dimensional integral (\ref{eq:vdWP2}) of Theorem 2 does not
present a well-posed inverse problem for the intermolecular potential
$\Phi$. However, the spherically averaged integral equation
(\ref{eq:lang2}) of Theorem 1 can be solved for the cell potential
$w(r)$ for a broad class of Langmuir curves $C(\beta)$ specified in
the following theorem. The proof is spread throughout
section~\ref{sec:general} of the main text.

\begin{theorem}
If the inverse Laplace transform $f(y)$ of $F(\beta)$ exists and is
nondecreasing and non-constant for $y > 0$, then there exist a unique
central-well solution ($r_o=0$) and infinitely many non-central-well
solutions ($0 < r_o < \infty$) to the inverse problem
(\ref{eq:langmuir}). If $f(y)$ is also continuous, then the
central-well solution is the only continuously differentiable
solution.
\end{theorem}

Finally, we state sufficient assumptions on $F(\beta)$ to guarantee
the assumed properties of $f(y)$.  In light of the necessary condition
that $F(\beta)$ be analytic the right half plane $\mbox{Re}\beta> c$,
the defining contour integral for $f(y)$
\begin{equation}
f(y) = \frac{1}{2\pi i} \int_{c^\prime - i\infty}^{c^\prime + i\infty}
e^{\beta y} F(\beta) d\beta
\end{equation}
must converge for any $c^\prime > c$.  By closing the contour in the
left half plane, it can be shown that a sufficient (but not necessary)
condition to ensure the assumed properties of $f(y)$ is that
$F(\beta)$ decay in the left half plane ($\lim_{\rho\rightarrow\infty}
|F(\rho e^{i\theta})| = 0$ for $\pi/2 \leq \theta \leq 3\pi/2$) and
have isolated singularities only on the negative real axis or at the
origin with positive real residues. The particular examples of
$F(\beta)$ considered in section~\ref{sec:deviations} satisfy these
conditions, but the weaker assumptions above regarding $f(y)$
suffice for the general derivation in section~\ref{sec:general}.

\newpage

\clearpage

\begin{table}
\begin{center}
\begin{minipage}[h]{5in}
\caption{Linear regressions of the experimental Langmuir constant
data~\protect\cite{sparks_thesis} for ethane and cyclopropane
clathrate-hydrates on the form $\log C = m\beta + b + \nu\log(\beta)$,
where $b = \log C_o$.
Errors reflect $63\%$ confidence intervals. The rows where $\nu=
0,1/2,$ or $-1/2$ correspond to two-parameter fits with $\nu$ held
constant.
\label{tab:fit} }
\end{minipage}
\\
\vspace{0.5in}
\begin{tabular}{|c|ccc|}
\hline
Guest Molecule & $m$ (kcal/mol) & $\log C_o$ (atm$^{-1}$) & $\nu$ \\
\hline
& & & \\
Ethane & 9.422 $\pm$ 0.043 & -14.561 $\pm$ 0.081 &   0 \\
       & 9.180 $\pm$ 0.044 & -14.419 $\pm$ 0.082 &  1/2 \\
       & 9.664 $\pm$ 0.043 & -14.703 $\pm$ 0.080 & -1/2 \\
& & & \\
       & 10.52 $\pm$ 0.85  & -15.2 $\pm$    0.50 & -2.3 $\pm$ 1.8 \\
& & & \\
Cyclopropane
      & 10.594 $\pm$ 0.012 & -15.474 $\pm$ 0.022 &   0 \\
      & 10.335 $\pm$ 0.011 & -15.302 $\pm$ 0.021 &  1/2 \\
      & 10.853 $\pm$ 0.012 & -15.646 $\pm$ 0.024 & -1/2 \\
& & & \\
      & 9.36 $\pm$ 0.47 & -14.66 $\pm$ 0.31 & 2.37 $\pm$ 0.90 \\
\hline
\end{tabular}
\end{center}
\end{table}

\end{document}